\def\Rsin{{R_{\text{S}_2\text{I}\text{N}}}}

\documentclass[aps,prappl,twocolumn,groupedaddress,superscriptaddress,floatfix,longbibliography]{revtex4-1}
\usepackage{epsfig}
\usepackage{multirow}
\usepackage{amsmath,amssymb,mathtools}
\usepackage{color}
\usepackage{graphicx}
\usepackage{dcolumn}
\usepackage{bm}
\usepackage{subfigure}
\usepackage[utf8]{inputenc}
\usepackage[T1]{fontenc}
\usepackage{tikz}

\graphicspath{{Figures/}}
\definecolor{AB-color}{RGB}{128,0,128}
\begin{document}

\title{Josephson threshold calorimeter}

\author{Claudio Guarcello}
\email{claudio.guarcello@nano.cnr.it}
\affiliation{NEST, Istituto Nanoscienze-CNR and Scuola Normale Superiore, Piazza San Silvestro 12, I-56127 Pisa, Italy}
\author{Alessandro Braggio}
\affiliation{NEST, Istituto Nanoscienze-CNR and Scuola Normale Superiore, Piazza San Silvestro 12, I-56127 Pisa, Italy}
\author{Paolo Solinas}
\affiliation{SPIN-CNR, Via Dodecaneso 33, 16146 Genova, Italy}
\author{Giovanni Piero Pepe}
\affiliation{Dipartimento di Fisica Ettore Pancini, Universit\'a degli Studi di Napoli Federico II, Napoli, Italy}
\affiliation{CNR-SPIN, Complesso Monte Sant'Angelo via Cinthia, I-80126 Napoli, Italy}
\author{Francesco Giazotto}
\affiliation{NEST, Istituto Nanoscienze-CNR and Scuola Normale Superiore, Piazza San Silvestro 12, I-56127 Pisa, Italy}

\begin{abstract}
We suggest a single-photon thermal detector based on the abrupt jump of the critical current of a temperature-biased tunnel Josephson junction formed by different superconductors, working in the dissipationless regime. The electrode with the lower critical temperature is used as a radiation sensing element, so it is thermally floating and is connected to an antenna/absorber. The warming up resulting from the absorption of a photon can induce a drastic measurable enhancement of the critical current of the junction. We propose a detection scheme based on a threshold mechanism for single-- or multi--photon detection. This Josephson threshold detector has indeed calorimetric capabilities being able to discriminate the energy of the incident photon. So, for the realistic setup that we discuss, our detector can efficiently work as a calorimeter for photons from the mid infrared, through the optical, into the ultraviolet, specifically, for photons with frequencies in the range $[30-9\times10^4]\;\text{THz}$. In the whole range of detectable frequencies, we obtain a resolving power significantly larger than one. In order to reveal the signal, we suggest the fast measurement of the Josephson kinetic inductance. Indeed, the photon-induced change in the critical current affects the Josephson kinetic inductance of the junction, which can be non-invasively read through an LC tank circuit, inductively coupled to the junction. Finally, this readout scheme shows remarkable multiplexing capabilities.
\end{abstract}

\maketitle

\section{Introduction}
\label{Sec00}\vskip-0.2cm

Superconducting electronics is nowadays efficiently employed for developing detectors for (single-) photon and particle calorimetry. These sensors are particularly appealing in view of their high detection efficiency, low energy threshold, and high energy resolution~\cite{Gia06}. Actually, the panorama of superconducting radiation detectors offers different approaches, that are quite different from each other and take advantages of peculiar features of the superconducting materials used for the devices. Among these, transition-edge sensors (TESs)~\cite{Ull15} are demonstrated high sensitivity for X-ray and $\gamma$-ray spectrometry. Instead, superconducting nanowire single photon detectors (SNSPDs)~\cite{Dau14} are specifically used for near-infrared single photon detection in quantum communication, according to their peculiar characteristics such as high speed, large detection efficiency, reduced timing jitter, and small dark count rates.

Low-temperature detectors provide a drastic thermal noise suppression and pave the way to quantum-mechanical phenomena at cryogenic temperatures~\cite{Gia06}. The suppression of thermal noise allows energy deposited by photons or particles to be detected with a high resolution. Nevertheless, the geometrical efficiency of superconducting sensors can be a limitation, and thus the progresses in this field concentrate on providing wide collecting areas, large absorption efficiency, and high energy acceptance.

Thermal detectors rely on the conversion of a temperature rise in a measurable variation of an electric signal. In particular, ``quantum'' calorimetry permits to detect single photons and measure their energy~\cite{Pek13,Don18,Bra18}. These detectors are characterized by shorter photon-induced energy deposition with fast internal equilibration times as compared to the characteristic thermal relaxation time. They require fast and ultra-sensitive thermometry~\cite{Gas15,Sai16,Wan18,Zgi18,Kar18} and cryogenic operating temperatures. Time-resolved photon detection finds applications in many research fields including quantum communication~\cite{Sil14}, security~\cite{Tak07}, and quantum thermodynamics~\cite{Pek13,Ber15,Pek15,Mar16,Mar18,Vis18}, and is nowadays recently receiving increasing attention.

\begin{figure}[b!!]
\centering
\includegraphics[width=\columnwidth]{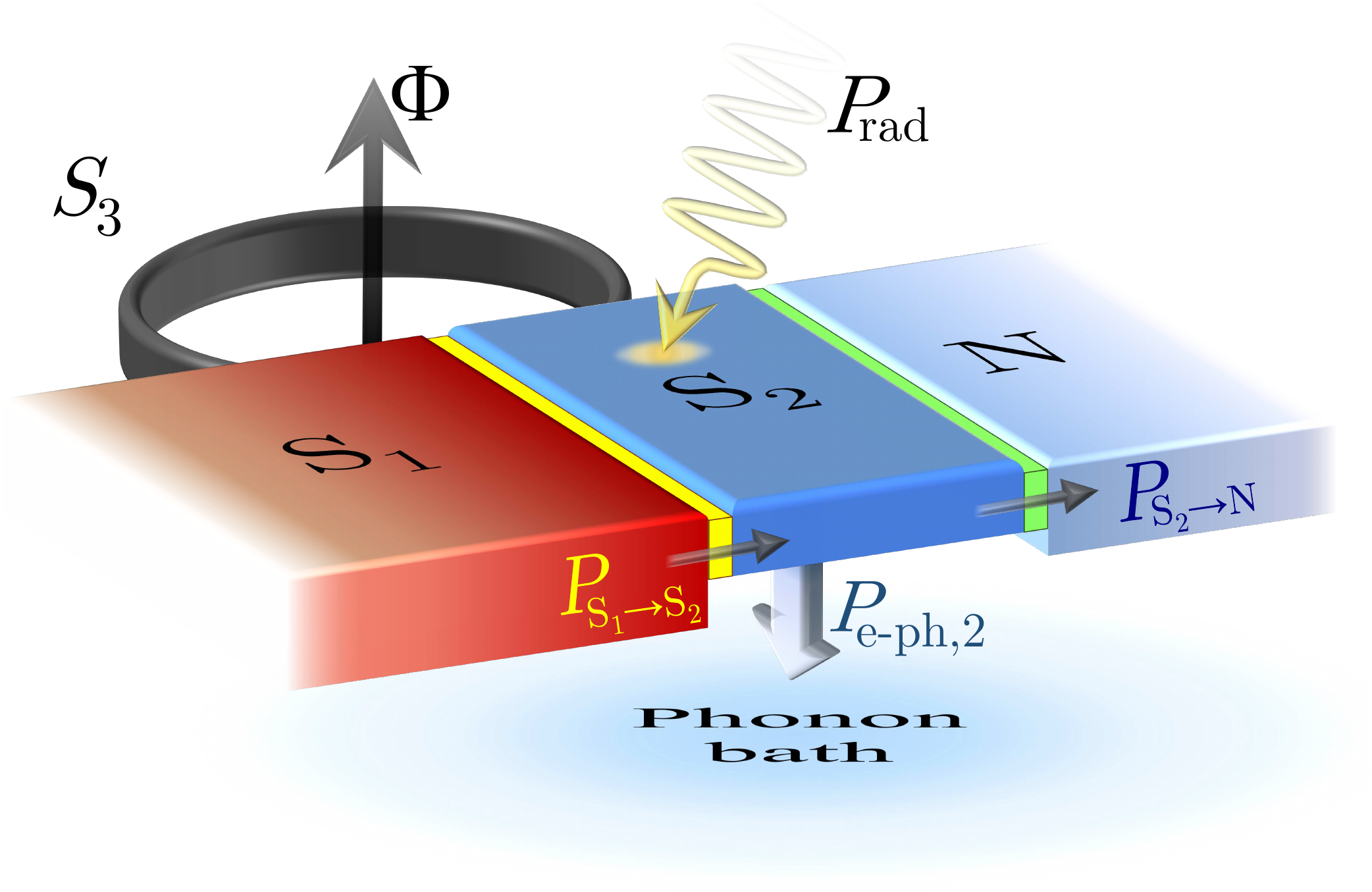}
\caption{Schematic illustration of a Josephson threshold calorimeter, that is a temperature-biased Josephson tunnel junction formed by the superconducting leads $S_1$ and $S_2$, with critical temperatures $T_{c_1}\neq T_{c_2}$, and residing at temperatures $T_1$ and $T_2$. The junction is enclosed in a superconducting ring pierced by a magnetic flux $\Phi$ which allows phase biasing of the weak link. The ring is supposed to be made by a third superconductor $S_3$ with energy gap $\Delta_3\gg\Delta_1,\Delta_2$, so to suppress the heat losses. The main thermal pathways in the JJ are also depicted. In detail, the phase-dependent thermal current from $S_1$ to $S_2$, $P_{\text{S}_1\to\text{S}_2}\left ( T_1,T_2,\varphi \right )$, the outgoing thermal currents from $S_2$ to the phonon bath and the cooling finger, $P_{e\text{-ph},2}\left ( T_2,T_{\text{bath}}\right )$ and $P_{\text{S}_2\to\text{N}}\left ( T_1,T_2,\varphi \right )$, respectively, and the photon-induced power diffusion in the absorber, $P_{\text{rad}}$, are also represented, for $T_1>T_2>T_{\text{bath}}$.}
\label{Fig01}
\end{figure}

Here, we suggest a dissipationless Josephson-based quantum calorimeter based on the peculiar sharp temperature dependence of the critical current in a Josephson tunnel junction formed by different superconductors, residing at different temperatures~\cite{GuaBra18} [see Fig.~\ref{Fig01}]. So, in our detector one needs to maintain a temperature gradient along the junction. By properly choosing the working temperatures of the junction, small photon-induced temperature increases could be revealed. Hereafter, we discuss a proposal based on a threshold mechanism, which gives us the possibility to develop a single photon detector with calorimetric capabilities. 
This Josephson threshold calorimeter (JTC) may be a new element in the panorama of recently designed superconducting devices that effectively take advantage of a thermal gradient imposed across the system. Specifically, we are dealing with solid-state applications in the realm of phase-coherent caloritronics~\cite{Gia06,MarSol14,ForGia17}, a research field that promises new ways to coherently master, store, and transport heat at the meso and nanoscopic scale. In fact, different kinds of temperature-based devices, such as heat interferometers~\cite{Gia12}, diffractors~\cite{Mar14,Gua16}, diodes~\cite{Mar15}, transistors~\cite{For16}, memories~\cite{Gua17,GuaSol17,GuaSol18}, logic elements~\cite{Pao18}, switches~\cite{Sot17}, routers~\cite{Tim18,Gua18}, and circulators~\cite{Hwa18} were recently conceived.

The paper is organized as follows. In Sec.~\ref{Sec01}, we introduce the working principles of a radiation detector based on a temperature-biased asymmetric Josephson junction (JJ), looking also the influence of the thermal fluctuations. In Sec.~\ref{Sec02}, we focus on the calorimeter detection modes of our device, discussing the idle state of the detector, proper figure of merits, i.e., the resolving power and frequency dynamical ranges, the time evolution of both the temperature and the critical current, and finally a readout scheme based on the measurement of the Josephson kinetic inductance, including a proposal of a multiplexing scheme. 
In Sec.~\ref{Sec03}, the conclusions are drawn.

\section{Detection operating principles}
\label{Sec01}\vskip-0.2cm

In this section we discuss the operating principles of a radiation detector based on the discontinuous thermal response of the critical current, $I_c$, of an asymmetric tunnel JJ~\cite{GuaBra18}. 
This behavior emerges by changing the electronic temperature of the electrodes of the junction. We are assuming that the lattice phonons in the electrodes are very well thermalized with the substrate, residing at the temperature $T_{\text{bath}}$, thanks to the vanishing Kapitza resistance between thin metallic films and the substrate at low temperatures~\cite{Wel94,Gia06}.

This peculiar behavior of $I_c$ derives from the matching of the superconducting gaps at specific temperatures. Essentially, we briefly recollect the main mechanisms discussed in detail in Ref.~\cite{GuaBra18}. Here we develop a careful investigation of the sensor design in order to achieve the best detection performances. The Josephson junction is asymmetric in the sense that the two superconducting leads $S_1$ and $S_2$, with energy gaps $\Delta_1$ and $\Delta_2$ and residing at temperatures $T_1$ and $T_2$, are made up of different BCS superconductors. We can define the asymmetry parameter 
\begin{equation}
r=\frac{T_{c_1}}{T_{c_2}}=\frac{\Delta_{0,1}}{\Delta_{0,2}},
\end{equation}
where $T_{c_j}$ is the critical temperature and $\Delta_{0,j}=1.764 k_BT_{c_j}$ is the zero-temperature superconducting BCS gap~\cite{Tin04} of the $j$-th superconductor (with $k_B$ being the Boltzmann constant). A film with the desirable critical temperature can be obtained by using a proximity-coupled bilayer, as it is usually done in TESs~\cite{Ull15}.
In a thermally biased Josephson tunnel junction the critical current reads~\cite{Gol04,Gia05,Tir08}
\begin{eqnarray}\label{IcT1T2Im}\nonumber
I_c\left ( T_1,T_2 \right )=&&\frac{1}{2eR}\Bigg |\underset{-\infty}{\overset{\infty}{\mathop \int }} \Big \{ F\left ( \varepsilon ,T_1 \right )\textup{Re}\left [\mathfrak{F}_1(\varepsilon,T_1 ) \right ]\textup{Im}\left [\mathfrak{F}_2(\varepsilon,T_2 ) \right ] \\&&+ F\left ( \varepsilon ,T_2 \right )\textup{Re}\left [\mathfrak{F}_2(\varepsilon,T_2 ) \right ]\textup{Im}\left [\mathfrak{F}_1(\varepsilon,T_1 ) \right ]\Big \} d\varepsilon\Bigg |.
\end{eqnarray}
Here, $R$ is the normal-state resistance of the junction, $e$ is the electron charge, $F\left ( \varepsilon ,T_j \right )=\tanh\left ( \varepsilon/2 k_B T_j \right )$, and
\begin{equation}\label{Green}
\mathfrak{F}_j(\varepsilon,T_j ) =\frac{\Delta_j \left ( T_j \right )}{\sqrt{\left ( \varepsilon +i\Gamma_j \right )^2-\Delta_j^2 \left ( T_j\right )}}
\end{equation}
is the anomalous Green's function of the $j$-th superconductor~\cite{Bar82}, with $\Gamma_j=\gamma_j\Delta_{0,j}$ being the Dynes parameter~\cite{Dyn78}. In this work, we set $\gamma_j =10^{-5}$, a value often used to describe realistic superconducting tunnel junctions~\cite{Gia06,Pek10}.

Fig.~\ref{Fig01} shows a possible experimental realization of the detector. The device also includes the thermal contact through a tunnel junction with a normal metal lead, $N$, having a large heat capacity and residing at the bath temperature. As we will discuss later, this ``cooling finger'' can play a predominant role in the thermal balance of the floating $S_2$ lead. It allows to control the working temperatures of the device, to extend the range of detectable photon frequencies, and to master the thermal response time of the device. 
Moreover, the resistance of this cooling finger, i.e., the resistance $\Rsin$ of the junction between $S_2$ and $N$, determines both the steepness of the abrupt variation of the critical current and the temperature at which it manifests itself. Finally, for low $\Rsin$'s, the cooling finger becomes the predominant thermal relaxation channel in the system and the thermalization process in $S_2$ becomes faster.

In Fig.~\ref{Fig01}, we also indicate how to control the phase difference $\varphi$ across the thermally-biased Josephson junction, which is enclosed, through clean contacts, within a superconducting ring pierced by a magnetic flux $\Phi$. In this way, the phase-biasing can be achieved via the external flux, since, when the ring inductance can be neglected, the phase-flux relation is given by $\varphi=2\pi\Phi/\Phi_0$~\cite{Cla04}. Here, $\Phi_0= h/2e\simeq2\times10^{-15}\; \textup{Wb}$ is the magnetic flux quantum, with $h$ being the Planck constant. Therefore, the phase drop across the junction can be varied within the whole phase space, i.e., $-\pi\leq\varphi\leq\pi$. The ring is supposed to be made by a third superconductor $S_3$ with energy gap $\Delta_3\gg\Delta_1,\Delta_2$. Under this conditions we neglect the heat transfer between $S_{1,2}$ and $S_3$ due to Andreev reflection heat mirroring effect~\cite{And64}.
Thus, only the temperature gradient between $S_1$ and $S_2$ is important for the junction. 

The JTC that we are conceiving is based on the critical current steeper behavior, which stems from the matching of the superconducting gaps. This phenomenon is better illustrated in the next section.

\subsection{The superconducting gap constraints}
\label{Sec01b}\vskip-0.2cm

\begin{figure}[t!!]
\centering
\includegraphics[width=\columnwidth]{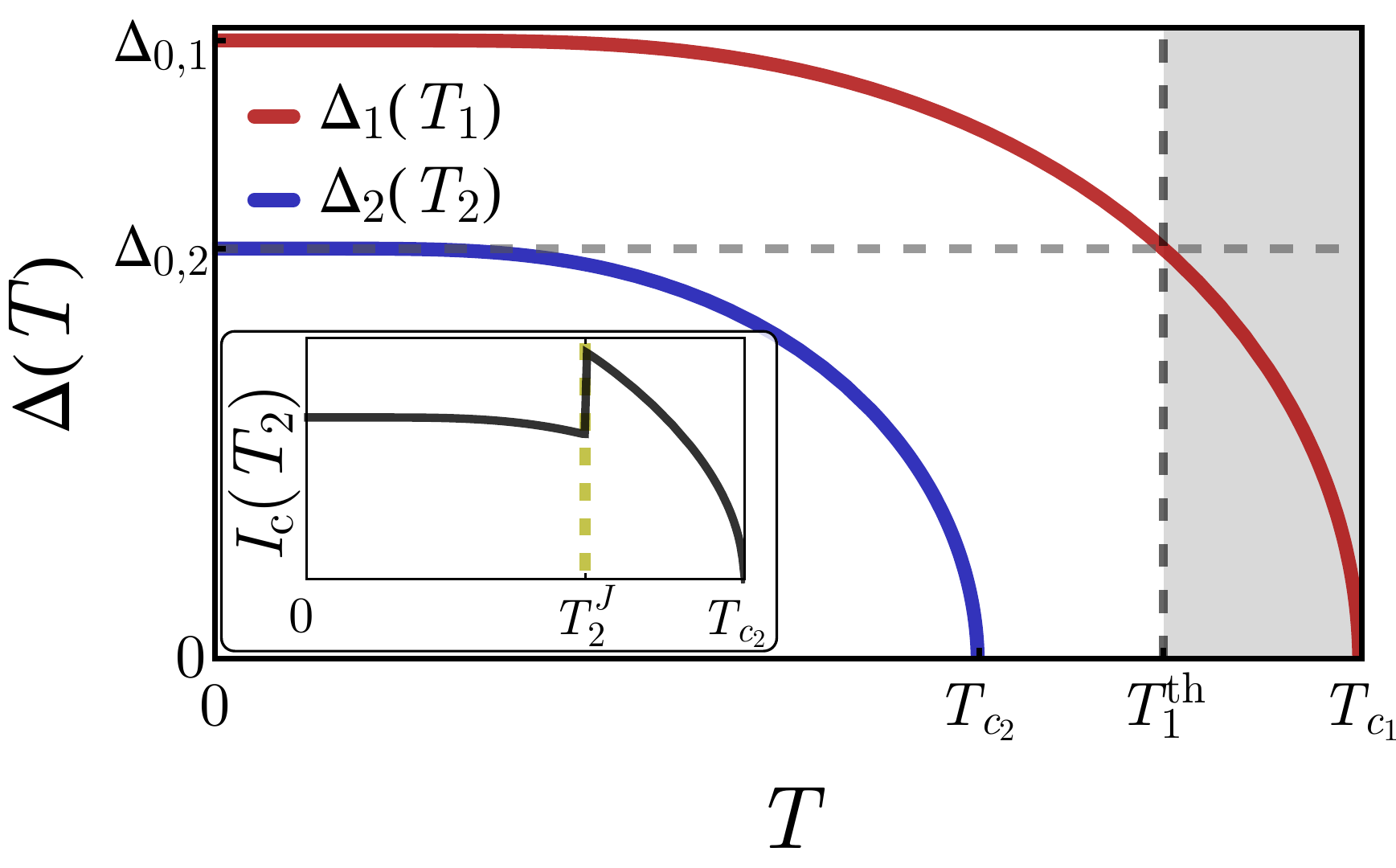}
\caption{Superconducting gaps as a function of the temperature. The shaded region highlights the range of $T_1$ values above the threshold $T_1^{\text{th}}$, that are suitable for our photonic detection scheme. In the inset: critical current profile showing a jump at a temperature $T_2=T_2^J$.}
\label{Fig02}
\end{figure}

With the aim to understand the conditions at which $I_c$ abruptly jumps, we show in Fig.~\ref{Fig02} the temperature dependence of the superconducting gaps, assuming $r>1$, i.e., $\Delta_{0,1}>\Delta_{0,2}$. The peculiar step-like behavior of the critical current, which is clearly shown in the inset of this figure where we show $I_c(T_2)$ at a fixed $T_1$, stems from the alignment of the singularities in the Green's functions $\mathfrak{F}_j$ at $\varepsilon=\Delta_j$, see Eq.~\eqref{Green}, when the superconducting gaps coincide~\cite{Bar82,GuaBra18}
\begin{equation}\label{GapEqs}
\Delta_1(T_1)=\Delta_2(T_2).
\end{equation}
We observe that for $r>1$ the superconducting gaps can be equal if and only if the temperature $T_1$ assumes a value higher than a threshold $T_1^{\text{th}}$, then for $T_1$ within the shaded region in Fig.~\ref{Fig02}. The value of $T_1^{\text{th}}$ depends on $r$ and can be calculated as the temperature at which $\Delta_1(T_1^{\text{th}})=\Delta_{0,2}$. Then, once the temperature $T_1\in[T_1^{\text{th}},T_{c_1}]$ is settled, the critical current jumps at a specific $T_2=T_2^J$ [see the dashed line in the inset of Fig.~\ref{Fig02}] at which $\Delta_2(T_2^J)=\Delta_1(T_1)$. Therefore, $T_2^J$ depends on the operating value of $T_1$, but it can in principle assume every value in the whole range $[0,T_{c_2}]$.

Following the previous discussion, it is worthwhile to choose the electrode with the lower $T_c$ as sensing element. In fact, this choice broadens the range of temperatures available as a result of a photonic energy absorption. Then, since we label the absorbing lead with $S_2$, this electrode has the lower critical temperature, that is $T_{c_2}<T_{c_1}$ and, consequently, $r>1$. This case corresponds to a sudden increase of $I_c$ at $T_2=T_2^J$~\cite{GuaBra18}. Conversely, for $ r <1 $ the critical current suddenly reduces at $T_2=T_2^J$. Thus, our choice for the sensing element is supported by the fact that a detection scheme in which a photonic event causes $I_c$ to increase provides a stronger Josephson coupling. Furthermore, an abrupt increase of the critical current will certainly result from an energy absorption in the floating lead, since a photon eventually absorbed in $S_1$ would not determine any steeper critical current variation. This configuration is also more robust against thermal fluctuations, in comparison to the case in which $I_c$ decreases.

We also note that by reducing the asymmetry of the junction, i.e., $r\to1$, the threshold temperature $T_1^{\text{th}}$ decreases. This could be beneficial in order to choose an operating point for $T_1$ not too close to the critical temperature. On the other hand, the height of $I_c$ jump, and therefore the sensor sensitivity, is expected to decrease when $r\to1$, vanishing for $r=1$.
Conversely, by increasing the asymmetry between the gaps, that is for $r\gg1$, we are suppressing one superconducting gap with respect to the other. In these cases, $T_1^{\text{th}}\to T_{c_1}$. 
This implies that the range of $T_1$ values suitable for the detection [see the grey region in Fig.~\ref{Fig02}] tends to reduce.
Furthermore, since in this case $I_c\to0$, we expect that the height of the $I_c$ jumps will tend to diminish. 

Therefore, one has to find an optimal value of the asymmetry parameter $r\gtrsim1$ which maximizes the sensitivity without requiring an operating temperature $T_1$ too close to $T_{c_1}$. In Ref.~\cite{GuaBra18} we found that the $I_c$ jump maximum is observed at $r\sim3$. Unfortunately, at this value the range of available $T_1$ values is too narrow, so we choose a slightly smaller value of $r=1.5$, in order to enlarge the dynamical range of the detector.
Then, assuming for the electrode $S_1$ the critical temperature $T_{c_1}=1.2\;\text{K}$, the critical temperature of $S_2$ has to be $T_{c_2}=0.8\;\text{K}$. This choice fixes the threshold value of $T_1$ to the value $T_1^{\text{th}}\simeq0.992\;\text{K}$.

Our detector is based on the capability to monitor the sudden increase of $I_c$ as a consequence of a rise of the temperature of the electrons in $S_2$ as the photon is absorbed.
In our scheme we also assume that the electrode $S_1$ has a large enough thermal capacity and is well thermally connected to heating probes (not depicted in Fig.~\ref{Fig01}). So, it resides at a stable temperature, which is weakly affected by changes of $T_2$. In this condition, we can neglect thermal fluctuations of $S_1$, treating it as a thermal reservoir.

To allow a photon to be detected, the absorbing element has to reside at a temperature, $T^0_2$, which is just below the temperature $T_2^J$ at which an $I_c$ jump occurs. Consequently, due to a photonic event the temperature $T_2$ can increase enough to exceed the threshold value, inducing an abrupt increment of the critical current. However, if $T^0_2$ is not close enough to $T_2^J$, lower energy photons may not trigger an appreciable $I_c$ enhancement. Conversely, if $T^0_2$ is too close to $T_2^J$, thermal fluctuations could induce undesired critical current jumps. Therefore, we need first of all to estimate the amplitude of the temperature fluctuations in $S_2$, in order to set a suitable detection threshold temperature, $\Delta T_2$, which is the distance between $T^0_2$ and $T_2^J$. Choosing an optimal detection threshold $\Delta T_2$ is essential for the proper functioning of the JTC and for minimizing the dark counts rate, i.e., the probability of false positive detections.

\subsection{The detection threshold temperature}
\label{Sec01d}\vskip-0.2cm

\begin{figure}[t!!]
\centering
\includegraphics[width=\columnwidth]{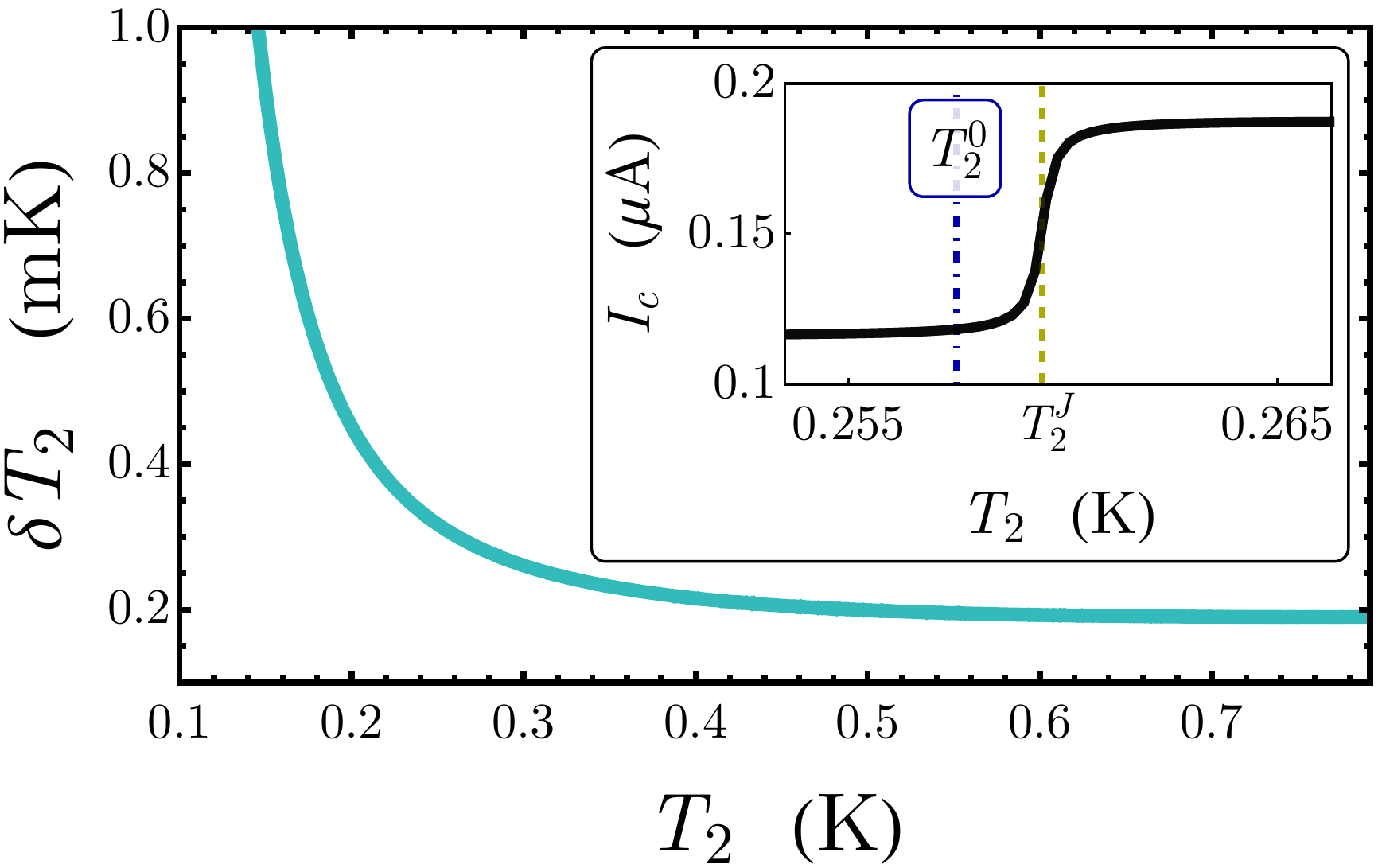}
\caption{Temperature fluctuation in $S_2$, see Eq.~\eqref{TemperatureFluctuation}, as a function of $T_2$. In the inset: Critical current, $I_c(T_1,T_2)$, as a function of $T_2$ near the jump, at $T_1=994.8\;\text{mK}$. The values of other parameters are: the critical temperatures are $T_{c_1}=1.2\;\text{K}$, $T_{c_2}=0.8\;\text{K}$, the JJ resistance is $R=1\;\text{k}\Omega$, and the $S_2$ volume is $V_2=1\;\mu\text{m}^3$. }
\label{Fig03}
\end{figure}

Thermodynamic fluctuations of $S_2$ can be estimated through the root-mean-square fluctuations in energy~\cite{Mos84,Chu92} $\delta E_2(T_2)=\sqrt{C_2(T_2)k_BT_2^2}$. Here, $C_2$ is the electronic heat capacity of $S_2$ 
\begin{equation}
C_2(T_2)=T_2 \,\partial \mathcal{S}_2/\partial T_2.
\label{heatcapacity}
\end{equation}
In this equation, $\mathcal{S}_2(T_2)$ is the electronic entropy of $S_2$, which is given by~\cite{Rab08,Sol16}
\begin{equation}
\mathcal{S}_2(T_2)=-4k_BN_{F,2}V_2\int_{-\infty}^{\infty}\!\!\!\!\!\!d\varepsilon f(\varepsilon,T_2) \ln[ f(\varepsilon,T_2)] \mathcal{N}_2(\varepsilon,T_2) ,
\label{Entropy}
\end{equation}
where $N_{F,2}$ is the density of states at the Fermi energy, $V_2$ is the volume of $S_2$, $f ( \varepsilon ,T_j )$ is the Fermi distribution function, and $\mathcal{N}_j\left ( \varepsilon ,T_j \right )=\left | \text{Re}\left [ \frac{ \varepsilon +i\Gamma_j}{\sqrt{(\varepsilon +i\Gamma_j) ^2-\Delta _j\left ( T_j \right )^2}} \right ] \right |$ is the smeared BCS density of states of the $j$-th superconducting lead. Then, we can evaluate the temperature fluctuation as~\cite{Wal17}
\begin{equation}
\delta T_2=\frac{\delta E_2(T_2)}{C_2(T_2)}=\sqrt{\frac{k_BT_2^2}{C_2(T_2)}}.
\label{TemperatureFluctuation}
\end{equation}
The behavior of thermal fluctuations, $\delta T_2$, in the absorbing lead $S_2$, setting $V_2=1\;\mu\text{m}^3$, as a function of $T_2$ is shown in Fig.~\ref{Fig03}. 
Here, the volume was selected on the base of the feasibility of the device. This volume will guarantee the possibility to connect the floating lead to both the electrode $S_1$ and the metallic cooling finger with realistic values of the resistances.
According to the exponential suppression of the electronic heat capacity in a superconductor at a low temperature~\cite{Abr75,DeG99}, we observe that $\delta T_2$ diverges for $T_2\to0$, and it monotonically decreases by increasing $T_2$ approaching the value $\delta T_2\simeq0.2\;\text{mK}$ at $T_2=T_{c_2}$.

With the aim to avoid unwanted transitions and to minimize the dark counts, one needs to set an idle temperature $T_2^0$ which is distant more than $\delta T_2$ from the threshold value $T_2^J$. 
The idle temperature $T_2^0$ has to be chosen so that its difference with $T_2^J$ [see the inset of Fig.\ref{Fig03}], hereafter defined as $\Delta T_2$, is larger with respect to the thermal fluctuations, but smaller than the temperature rise following the absorption of a photon.

For a detector with a sensing element volume $V_2=1\;\mu\text{m}^3$, we set a detection threshold equal to $\Delta T_2=2\;\text{mK}$, since it would be roughly 10 times more than the $\delta T_2$ values shown in Fig.~\ref{Fig03}, in the range of operating $T_2$ temperatures that we are going to consider. This choice allows to avoid unwanted transitions eventually triggered by thermodynamic fluctuations. Additionally, as we will discuss in the next section, this value of $\Delta T_2$ still permits to detect photons in a wide range of frequencies, since its value affects the minimum detectable photon frequency, $\nu_{\text{min}}$. 
Thus, once we have chosen the value of the detection threshold, to distinguish a photonic event we require a photon-induced temperature increase at least higher than $\Delta T_2$. However, first of all, we need to calculate the operating temperatures $T^0_1$ and $T^0_2$ by numerically solving a heat balance stationary equation. Then, in the next section we will thoroughly describe all predominant heat exchanges involved in the thermal model.

\subsection{The heat exchanges}
\label{Sec01a}\vskip-0.2cm

In Fig.~\ref{Fig01}, the thermal pathways in our device for $T_1>T_2>T_{\text{bath}}$ are depicted.
Once a thermal gradient along the system is imposed, a phase-dependent heat flux, $P_{\text{S}_1\to\text{S}_2}\left ( T_1,T_2,\varphi \right )$, flows through the junction from $S_1$ to $S_2$~\cite{Mak65}. The temperature $T_2$ will depend also on the outgoing thermal powers $P_{e\text{-ph,2}}\left ( T_2,T_{\text{bath}}\right )$ and $P_{\text{S}_2\to\text{N}}\left ( T_2,T_{\text{bath}}\right )$ relaxed towards the phonon bath and the metallic cooling finger, respectively.

The stationary phase-dependent total thermal power flowing from $S_1$ to $S_2$ reads
\begin{equation}\label{Pt}
P_{\text{S}_1\to\text{S}_2}( T_1,T_2,\varphi)=P_{\text{qp}}( T_1,T_2)-\cos\varphi \;P_{\cos}( T_1,T_2).
\end{equation}
In the adiabatic regime, that is the case of a negligible voltage drop with respect to the relevant energy scales in the system, $eV\ll \text{min} \left \{ k_BT_1, k_BT_2, \Delta_1(T_1), \Delta_2(T_2) \right \}$, the terms in Eq.~\eqref{Pt} read~\cite{Gol13}
\begin{eqnarray}\label{Pqp}\nonumber
P_{\text{qp}}(T_1,T_2)&=&\frac{1}{e^2R}\int_{-\infty}^{\infty} d\varepsilon\;\varepsilon \mathcal{N}_1 ( \varepsilon ,T_1 )\mathcal{N}_2 ( \varepsilon ,T_2 )\\
&\times& [ f ( \varepsilon ,T_1 ) -f ( \varepsilon ,T_2 ) ],
\end{eqnarray}
\begin{eqnarray}\nonumber
\label{Pcos}
P_{\cos}( T_1,T_2 )&=&\frac{1}{e^2R}\int_{-\infty}^{\infty} d\varepsilon \mathcal{N}_1 ( \varepsilon ,T_1 )\mathcal{N}_2( \varepsilon ,T_2 )\\
&\times&\frac{\Delta_1(T_1)\Delta_2(T_2)}{\varepsilon}[ f ( \varepsilon ,T_1 ) -f ( \varepsilon ,T_2 ) ].
\end{eqnarray}

Heuristically, Eq.~\eqref{Pt} derives from processes involving quasiparticles and coherent factors in tunnelling through a JJ, as predicted by Maki and Griffin~\cite{Mak65}. $P_{\text{qp}}$ is the heat current carried by quasiparticles, i.e., an incoherent power flow through the junction from the hot to the cold electrode~\cite{Mak65,Gia06}. Instead, the term $P_{\cos}$ determines the phase-dependent part of the heat transport originating from the energy-carrying tunneling processes involving recombination/destruction of Cooper pairs on both sides of the junction~\cite{Note2}.

The thermal current flowing through the $\text{S}_2\text{I}\text{N}$ junction is
\begin{eqnarray}\label{PSIN}\nonumber
P_{\text{S}_2\to\text{N}}(T_2,T_{\text{bath}})&=&\frac{1}{e^2\Rsin}\int_{-\infty}^{\infty} d\varepsilon\;\varepsilon \mathcal{N}_2 ( \varepsilon ,T_2 )\\
&\times& [ f ( \varepsilon ,T_2 ) -f ( \varepsilon ,T_{\text{bath}} ) ].
\end{eqnarray}
For low $\Rsin$, the cooling finger can become the predominant thermal relaxation channel. Instead, for high $\Rsin$, the heat relaxed through the cooling finger competes with the heat exchanged between electrons and phonons, the latter being thermalized at $T_{\text{bath}}$.

The term $P_{e\text{-ph},2}$ in Eqs.~\eqref{StatBalanceEqs} and~\eqref{DynamicBalanceEqs} represents the energy exchange between electrons and phonons in $S_2$ and reads~\cite{Kop01,Pek09}
\begin{eqnarray}\label{Qe-ph}\nonumber
P_{e\text{-ph},2}&=&\frac{-\Sigma_2V_2}{96\zeta(5)k_B^5}\int_{-\infty }^{\infty}dEE\int_{-\infty }^{\infty}d\varepsilon \varepsilon^2\textup{sign}(\varepsilon)M^2_{_{E,E+\varepsilon}}\\\nonumber
&\times& \Bigg\{ \coth\left ( \frac{\varepsilon }{2k_BT_{\text{bath}}}\right ) \Big [ F(E,T_2)-F(E+\varepsilon,T_2) \Big ]\\
&-&F(E,T_2)F(E+\varepsilon,T_2)+1 \Bigg\},
\end{eqnarray}
where $M^2_{E,{E}'}=\mathcal{N}_2(E,T_2)\mathcal{N}_2({E}',T_2)\left [ 1-\Delta_2 ^2(T_2)/(E{E}') \right ]$, $\Sigma_2$ is the electron-phonon coupling constant, $V_2$ is the volume of $S_2$, and $\zeta$ is the Riemann zeta function. 

In the heat exchange analysis we neglect any contribution from the photonic radiative channel, as discussed in Ref.~\cite{Bos16}. Indeed, the superconductors $S_1$ and $S_2$ are electrically connected via the JJ and the $S_3$ superconducting ring. In this configuration one could speculate if a pure radiative contribution has to be considered too. Anyway, this contribution can be neglected for two main reasons. Firstly, this contribution is several orders of magnitude lower than the quasiparticle galvanic contribution. Then, the radiative term to be effective requires an efficient impedance matching, as a result of the circuital configuration. In our setup, the impedance matching is not satisfied, since both the JJ and the superconducting ring have quite different lumped element schematization leading to a strong impedance mismatch for photonic transport. 

\begin{figure*}[t!!]
\centering
\includegraphics[width=2\columnwidth]{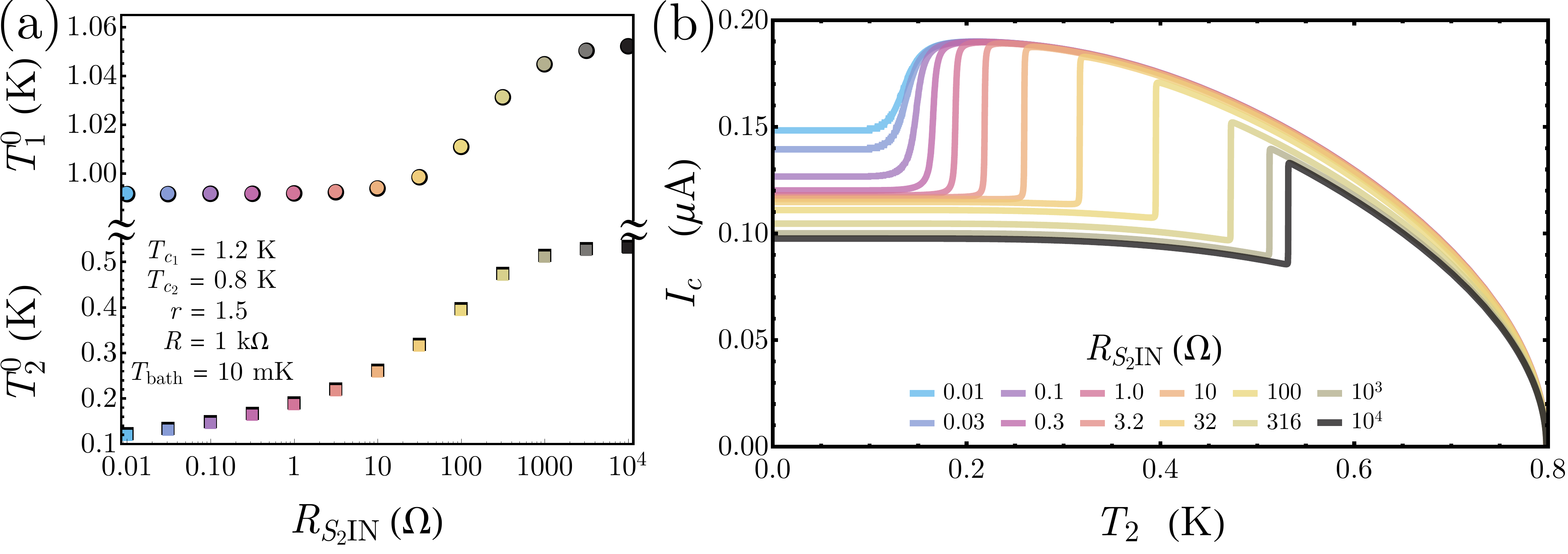}
\caption{(a) Idle temperatures, $T_1^0$ and $T_2^0$, as a function of the resistance $\Rsin$. (b) Critical current $I_c(T_1,T_2)$ as a function of the temperature $T_2$. The curves are obtained by imposing $T_1$ equal to the idle temperatures $T_1^0$ calculated in panel (a) changing $\Rsin$. The values of the other parameters are: $T_{c_1}=1.2\;\text{K}$, $T_{c_2}=0.8\;\text{K}$, $R=1\;\text{k}\Omega$, $T_{\text{bath}}=10\;\text{mK}$, $\varphi=0$, $\Sigma_2=0.3\times10^9\;\textup{W}\textup{m}^{-3}\textup{K}^{-5}$, and $N_{F,2}=10^{47}\;\text{J}^{-1}\text{m}^{-3}$. }
\label{Fig04}
\end{figure*}

\section{Calorimeter}
\label{Sec02}\vskip-0.2cm

Depending on the characteristic timescales of the process, the JJ can operate as a bolometer or as a calorimeter. Specifically, the detector works as a bolometer if the mean time between the arrival of incident photons is much shorter than the characteristic thermal relaxation time of the device. In the opposite regime, as the photonic arrival time exceeds the thermal relaxation time, the detector operates as a calorimeter. Therefore, a bolometer measures the total amount of radiation incident on an active area, whereas a calorimeter measures the energy of each single-photon absorption event~\cite{Kra96,Eis11,Ber13}.

Hereafter we propose a calorimetric detection scheme based on a threshold mechanism for single-- or multi--photon detection that takes advantages of the discussed very sharp variation of the critical current. 
During the design stage of this detector we realized the possibility of developing a different type of detection scheme, suitable for bolometric sensing, which is still based on the same physical mechanism. It will be thus considered elsewhere.

\subsection{The operating temperatures}
\label{Sec01c}\vskip-0.2cm

In order to evaluate the thermal evolution following the absorption of a photon in $S_2$, we firstly need to determine the stationary operating temperatures in the absence of photonic excitation, $T_1^0$ and $T_2^0$, suitable for our detection scheme. 
So, once we have chosen the value of the detection threshold $\Delta T_2$, we need to identify the stationary operating temperatures $T_1^0$ and $T_2^0$ yielding an idle state of the detector close enough to a jump in the critical current, that is the temperatures at which
\begin{equation}\label{GapEqsMod}
\Delta_1(T_1^0)=\Delta_2(T_2^0+\Delta T_2).
\end{equation}
Here, we are labelling these temperatures with $T_1^0$ and $T_2^0$, since in next sections we will use them as the initial values for the time-dependent analysis. Now, in order to tune the detector at a specific working point $(T_1^0,T_2^0)$, the temperature $T_2^0$ of the floating lead has to be obtained by numerically solving a steady-state heat balance equation for $S_2$,
\begin{equation}\label{StatBalanceEqs}
P_{\text{S}_1\to\text{S}_2}(T_1^0,T_2^0,\varphi)-P_{\text{e-ph,2}}(T_2^0,T_{\text{bath}})-P_{\text{S}_2\to\text{N}}(T_2^0,T_{\text{bath}})=0,
\end{equation}
in the presence of the gap constraint Eq.~\eqref{GapEqsMod} and for fixed values of $T_{\text{bath}}$ and $\varphi$. 
The key quantities for setting properly the idle temperatures of the detector are the volume $V_2$ and the resistances $R$ and $\Rsin$ of the junctions. The steady temperature-bias established between the electrodes strongly depends on these quantities, since the higher $R$ and $\Rsin$, the narrower the thermal channels towards and from $S_2$, since $P_{\text{S}_1\to\text{S}_2}\propto R^{-1}$ and $P_{\text{S}_2\to\text{N}}\propto \Rsin^{-1}$. Whereas the larger $V_2$, the better the thermal coupling of the absorber with the phonon bath, since $P_{\text{e-ph,2}}\propto V_2$. 

The selection of a specific device configuration (i.e., the values of $R$, $\Rsin$, and $V_2$) is a crucial point to best set the operation mode and the performances of the detector. We use the resistance of the cooling finger as a knob to set the operating point of the device. This means, at this stage, to properly choose the values or $R$ and $V_2$ and to study how $\Rsin$ affects the operations of the detector. In fact, in order to reduce the operative value of $T^0_2$ one can make the thermal contact with the normal lead more transparent, by lowering the resistance $\Rsin$. Reducing the temperature $T^0_2$ can be advantageous since it leads both to the enhancement of the amplitude of the $I_c$ jumps, as it is discussed in Ref.~\cite{GuaBra18}, and to a larger range of detectable photonic frequencies. 

We observe that the higher the JJ normal-state resistance $R$, the lower the maximum value of the critical current, see Eq.~\eqref{IcT1T2Im}. To choose the value of $R$, we require that the Josephson energy $E_J=\frac{\Phi_0}{2\pi}I_c$ is much larger than the thermal energy $k_BT$. For instance, by assuming the value $R=1\;\text{k}\Omega$ we obtain a maximum critical current of $\sim0.2\;\mu\text{A}$, corresponding to a Josephson energy in units of $k_B$ of $\sim 5.5\;\text{K}$, a value well above the working temperatures that we will discuss later.  

Then, one can assume to keep fixed the resistance $R=1\;\text{k}\Omega$ and the volume $V_2=1\;\mu\text{m}^3$, and to find for each value of $\Rsin$ the idle temperatures $T_1^0$ and $T_2^0$. 
This is displayed in Fig.~\ref{Fig04}(a), at a fixed $R=1\;\text{k}\Omega$ and $V_2=1\;\mu\text{m}^3$, for $T_{\text{bath}}=0.01\;\text{K}$ and $\varphi=0$.

The thermal bath is kept at a low temperature, reducing the strength of the phononic thermalization channel and to enhance at the same time the effectiveness of the metallic cooling finger. The need for an accurate phase regulation, which is performed through the external flux $\Phi$, is significantly relaxed, since we observed that at low values of $\Rsin$ the idle temperatures are only slightly affected by a modifications in $\varphi$ (not shown). 

At low $\Rsin$'s, $T_1^0$ and $T_2^0$ tend to the threshold value $T_1^{\text{th}}$ and the bath temperature, respectively [see Fig.~\ref{Fig04}(a)]. By increasing the resistance of the $\text{S}_2\text{I}\text{N}$ junction the idle temperatures monotonically rise. In the high resistances limit, i.e., $\Rsin\gtrsim 1\;\text{k}\Omega$, both temperatures reach a plateau, since this relaxation channel is too weak and does not affect the idle state of the device. 

Therefore, each value of $\Rsin$ corresponds to specific operating temperatures $(T_1^0,T_2^0)$. This means that a given $\Rsin$ results in a specific critical current profile, $I_c(T_1^0,T_2)$, as it is shown in Fig.~\ref{Fig04}(b). In this figure we present the behavior of $I_c$ as a function of $T_2$. These curves are obtained by imposing the temperatures $T_1$ coinciding with the idle temperatures $T_1^0$ shown in Fig.~\ref{Fig04}(a) by changing $\Rsin$. By increasing the resistance, since $T_1^0$ increases, we observe that the $I_c$ jump shifts towards high $T_2$ values, becoming lower and steeper. Conversely, for low $\Rsin$'s the $I_c$ jump tends to become smoother. In this regard, we note that the sharpness of the jump depends usually on the inverse of the Dynes parameter, $\Gamma_j^{-1}$, as it is discussed in Ref.~\cite{GuaBra18}. Anyway, here we are assuming to keep constant the value of the Dynes parameter, so that the smoothing of the $I_c$ jump in Fig.~\ref{Fig04}(b) can be ascribed instead to the peculiar temperature dependence of the gap.

In fact, we notice that the jump in the $I_c$ profile tends to become smoother at low $\Rsin$. In this case, the jump is located at low temperatures, roughly when the temperature $T_2^J$ is less than $T_2\lesssim0.4\;T_c$. At these temperatures, the superconducting gap is roughly flat [see Fig.~\ref{Fig02}], and, in such a case, the variation of $T_2$ reflects weakly on the gap behavior, inducing a smoother $I_c$ jump.

In summary, the selection of $\Rsin$ makes it possible to engineer both the position and the sharpness of the jump in the critical current profile. A small value of $\Rsin$ ensures a low working temperature, and therefore a larger dynamical range of the detector, but it corresponds to a smooth $I_c$ variation at the jump. Conversely, for the threshold calorimetric mechanism that we are suggesting, it is more convenient to have a sharp $I_c$ transition. Nonetheless, by increasing further $\Rsin$ the height of the $I_c$ jump tends to reduce. Then, we settle the value $\Rsin=10\;\Omega$ which guarantees a steep and sufficiently high $I_c$ jump. This resistance gives an idle temperature $T_2^0=0.258\;\text{K}$. 

Finally, we verify that the temperatures $T_1^0$ and $T_2^0$, obtained by solving Eqs.~\eqref{GapEqsMod} and~\eqref{StatBalanceEqs}, at $\Rsin=10\;\Omega$ as shown in Fig.~\ref{Fig04}(a), give a critical current value just below a $I_c$ jump. 
In the inset of Fig.~\ref{Fig03} we show the critical current profile as a function of the temperature $T_2$, of a JJ with $T_{c_1}=1.2\;\text{K}$, $T_{c_2}=0.8\;\text{K}$, and $R=1\;\text{k}\Omega$, at a fixed temperature $T_1=994.8\;\text{mK}$. This temperature corresponds to the $T^0_1$ value obtained for $\Rsin=10\;\Omega$ [see Fig.~\ref{Fig04}(a)]. The dot-dashed line indicates the temperature $T_2^0=0.258\;\text{K}$, and it allows to clearly highlight the closeness of the idle temperature to the temperature $T_2^J$ (see the dashed line) at which $I_c$ jumps. 

Now, it is convenient to estimate through simple arguments, namely, without addressing yet the full time evolution of the temperature $T_2$, the dynamical ranges of the calorimeter, that is the photon-induced temperature rise and the range of detectable photonic frequencies.

\subsection{Dynamical range of the calorimeter}
\label{Sec02aa}\vskip-0.2cm

\begin{figure}[t]
\centering
\includegraphics[width=\columnwidth]{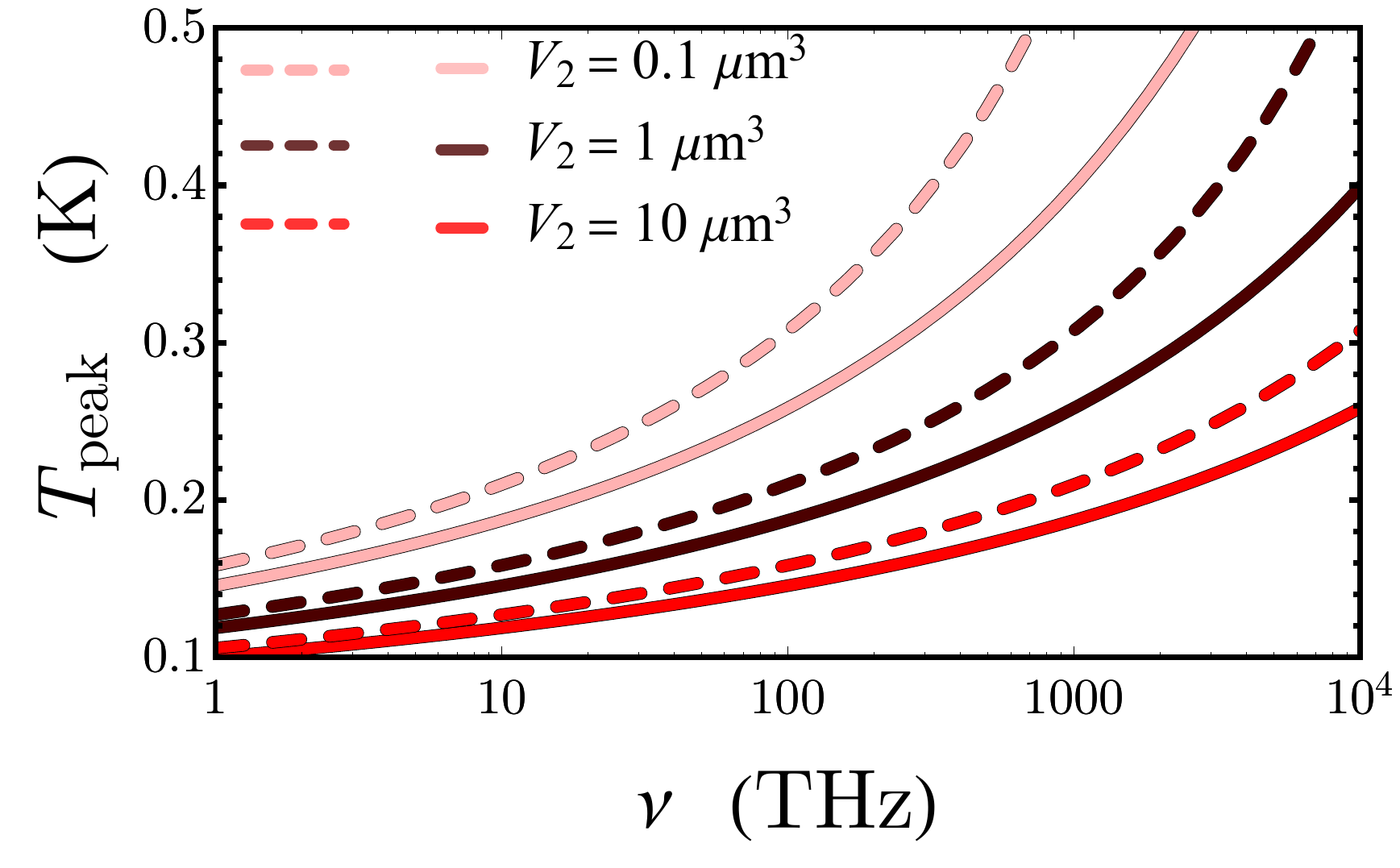}
\caption{Peak temperature, $T_{\text{peak}}$, as a function of the photon frequency, $\nu$, for a few values of the volume, $V_2$. Solid lines are obtained numerically, see Eq.~\eqref{TpeakRelation}, at $T_2^0=0$, whereas the dashed lines are calculated analytically in the low temperature limit through Eq.~\eqref{Tpeak0}.}
\label{Fig05}
\end{figure}

In the following, we assume that the system is always in a quasi-equilibrium state and that all the photon energy is transformed into internal energy of the electron.
This is a simplified situation since the dynamics of the superconducting lead is much more complicated. However, as we will discuss in Sec.~\ref{Sec02d}, for the time scale and the parameters used, these are good approximations.
So, we assume in a first approximation the full conversion of the photon energy to internal energy of electrons, i.e., we neglect any energy losses during the initial thermalization process following a photonic event. Indeed, we can thermodynamically estimate, in a very simple and direct manner, the temperature rise, $T_{\text{peak}}$, of the absorber. Specifically, $T_{\text{peak}}$ can be computed by equating the integrated internal energy of $S_2$ with the photon energy, that is
\begin{equation}
\int_{T_2^0}^{T_{\text{peak}}}C_2(T)dT=h\nu.
\label{TpeakRelation}
\end{equation}
We point out that in Sec.~\ref{Sec02d} we will calculate the full thermal evolution following a photon absorption including all heat exchange terms, accounting for all losses and heating processes in $S_2$. 

In the low temperature limit, $T\ll T_c$, we can give an analytic expression of $T_{\text{peak}}$. In fact, in this temperature regime the entropy of a BCS superconductor behaves exponentially with the temperature as~\cite{Abr75,DeG99}
\begin{equation}
S(T)\simeq k_BVN_F\Delta_0\sqrt{2\pi}\sqrt{\frac{\Delta_0}{k_BT}}e^{-\frac{\Delta_0}{k_BT}},
\label{Entropy0}
\end{equation}
from which the heat capacity at the leading order reads
\begin{equation}
C(T)\simeq k_BVN_F\Delta_0\sqrt{2\pi}\left (\frac{\Delta_0}{k_BT} \right )^{\frac{3}{2}}e^{-\frac{\Delta_0}{k_BT}}.
\label{HeatCapacity0}
\end{equation}

By inserting Eq.~\eqref{HeatCapacity0} in Eq.~\eqref{TpeakRelation}, and using the following integral identity $$\int_0^{a} dx\frac{e^{-\frac{1}{x}}}{x^{\frac{3}{2}}}=\sqrt{\pi}\left [1-\text{erf}\left ( \frac{1}{\sqrt{a}} \right ) \right ],$$ where $\text{erf}(x)$ is the error function, one finds
\begin{equation}
 h\nu\simeq \varepsilon\, V_2\left [ \text{erf}\left ( \sqrt{\frac{\Delta_{0,2}}{k_BT_2^0}} \right )-\text{erf}\left ( \sqrt{\frac{\Delta_{0,2}}{k_BT_{\text{peak}}}} \right ) \right ],
\label{Energy0-bis}
\end{equation}
with $\varepsilon=\sqrt{2}\pi N_{F,2}\Delta_{0,2}^2$. In the low temperature regime, i.e., $T_2^0\ll T_{c_2}$, where $\text{erf}\left ( \sqrt{\frac{\Delta_{0,2}}{k_BT_2^0}} \right )\to1$, we can determine from Eq.~\eqref{Energy0-bis} the analytical dependence of the peak temperature on both the volume $V_2$ and the photon frequency $\nu$, that reads
\begin{equation}
T_{\text{peak}}\simeq\frac{\Delta_{0,2}}{ k_B \left [ \text{erf}^{-1}\left ( 1 -\frac{h\nu}{\varepsilon V_2}\right ) \right ] ^2}.
\label{Tpeak0}
\end{equation}

Fig.~\ref{Fig05} shows a comparison between the peak temperatures calculated analytically in the low temperature limit through Eq.~\eqref{Tpeak0} (dashed lines) and computed numerically through Eq.~\eqref{TpeakRelation} (solid lines) taking the full temperature dependence of the entropy, see Eq~\eqref{Entropy}, and assuming $T_2^0=0$. We observe that the analytic estimate of $T_{\text{peak}}$ is systematically higher than the numerical one. This is because at high temperatures Eq.~\eqref{HeatCapacity0} underestimates the full heat capacity. Nevertheless, the numerical calculation converges to the analytic result in the low frequency case, in other words the regime of low temperatures where Eq.~\eqref{Entropy0} holds.

\begin{figure}[t!!]
\centering
\includegraphics[width=\columnwidth]{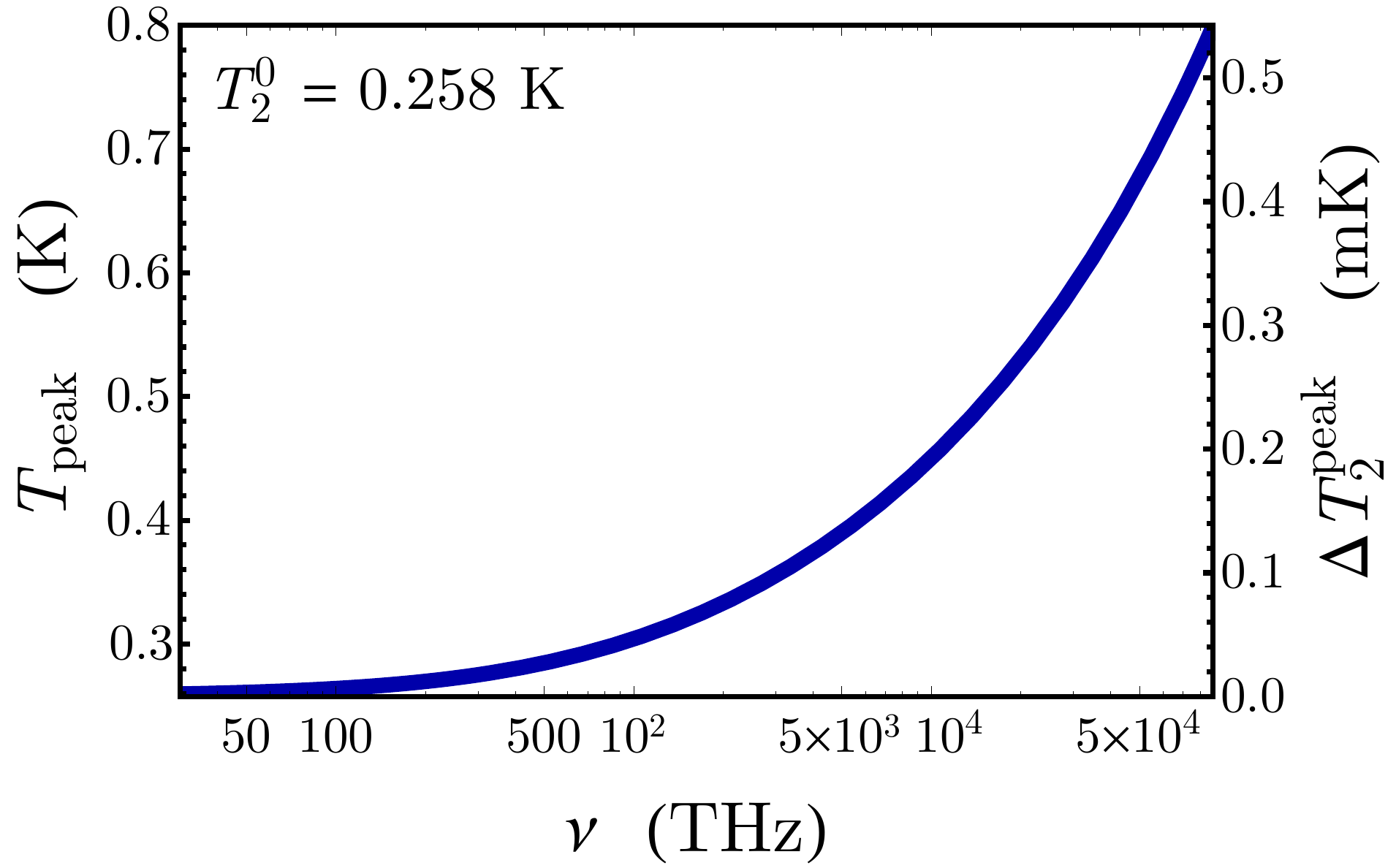}
\caption{Maximum temperature, $T_{\text{peak}}$, (left axis) and $T_2$ rise, $\Delta T^{\text{peak}}_2=T_{\text{peak}}-T_2^0$, (right axis) as a function of the photon frequency, $\nu$, at the base temperature $T_2^0=0.258\;\text{K}$.}
\label{Fig06}
\end{figure}

The Eq.~\eqref{TpeakRelation} allows also to directly estimate the detection frequency range of the device.  In Fig.~\ref{Fig06} we plot the behavior of $T_{\text{peak}}(\nu)$ (left vertical axis), at a base temperature equal to $T_2^0=0.258\;\text{K}$, which is the working temperature obtained by choosing $\Rsin=10\;\Omega$. On the right vertical axis of this panel the photon-induced temperature rise, $\Delta T^{\text{peak}}_2=T_{\text{peak}}-T_2^0$, is shown. 

Moreover, from the previous discussion we can derive both the the maximum photon energy $\nu_{\text{max}}$, that is the energy high enough to induce a superconducting-to-normal phase transition [$T_{\text{peak}}(\nu_{\text{max}})=T_{c_2}$ in Eq.~\eqref{TpeakRelation}], and the minimum detectable photon energy $\nu_{\text{min}}$, that is the energy giving a temperature rise just higher than the detection threshold [$T_{\text{peak}}(\nu_{\text{min}})=T_2^0+\Delta T_2$ in Eq.~\eqref{TpeakRelation}]. 
For instance, at $\Rsin=10\;\Omega$ [see Fig.~\ref{Fig06}], the proposed setup could effectively detect photons with frequencies up to $\nu_{\text{max}}\simeq9\times10^4\;\text{THz}$. Yet, in this case to make the photon-induced heating $\Delta T^{\text{peak}}_2$ at least higher than the detection threshold temperature $\Delta T_2=2\;\text{mK}$, it is necessary a photon with frequency above the threshold value $\nu_{\text{min}}\simeq30\;\text{THz}$.

\begin{figure}[t!!]
\centering
\includegraphics[width=\columnwidth]{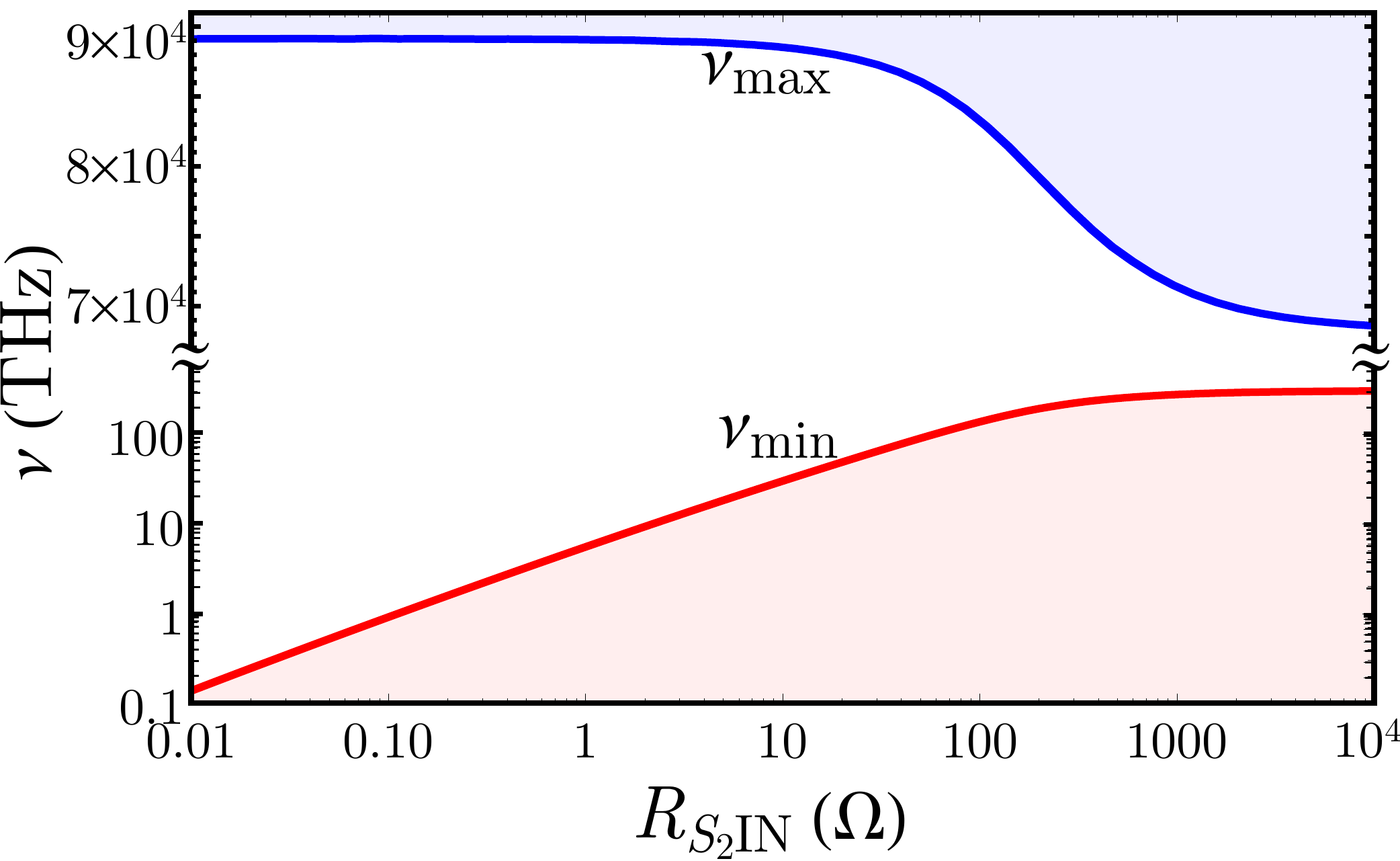}
\caption{Threshold frequencies $\nu_{\text{max}}$ and $\nu_{\text{min}}$ as a function of the resistance $\Rsin$.}
\label{Fig07}
\end{figure}

Since the idle temperature $T_2^0$ depends on $\Rsin$ [see Fig.~\ref{Fig04}(a)], the detection frequency range of the device changes by modifying the value of $\Rsin$. Fig.~\ref{Fig07} shows the behaviour of the threshold frequencies $\nu_{\text{max}}$ and $\nu_{\text{min}}$ as a function of $\Rsin$. 
Shaded regions indicate forbidden frequency values, since larger than $\nu_{\text{max}}$ or smaller than $\nu_{\text{min}}$.
We observe that the range of permitted frequencies reduces by increasing $\Rsin$. So, the choice of $\Rsin$ significantly affects not only the working temperature, but also the detection frequency range of the sensor.

The preliminary analysis performed here through Eq.~\eqref{TpeakRelation} allows checking if the initial temperature rise, evaluated in Sec.~\ref{Sec02d} dealing with the full thermal evolution in quasi-equilibrium conditions, is consistent with the thermodynamics of the process.

Finally, we observe that a photon can be absorbed if its energy is $h\nu\gtrsim2\Delta_2(T_2)$. For $T_{c_2}=0.8\;\text{K}$, we obtain $\nu\gtrsim50\;\text{GHz}$ at $T_2=0.258\;\text{K}$.

\subsection{Resolving power}
\label{Sec02c}\vskip-0.2cm

\begin{figure}[t!!]\label{ResolvingPower}
\centering
\includegraphics[width=\columnwidth]{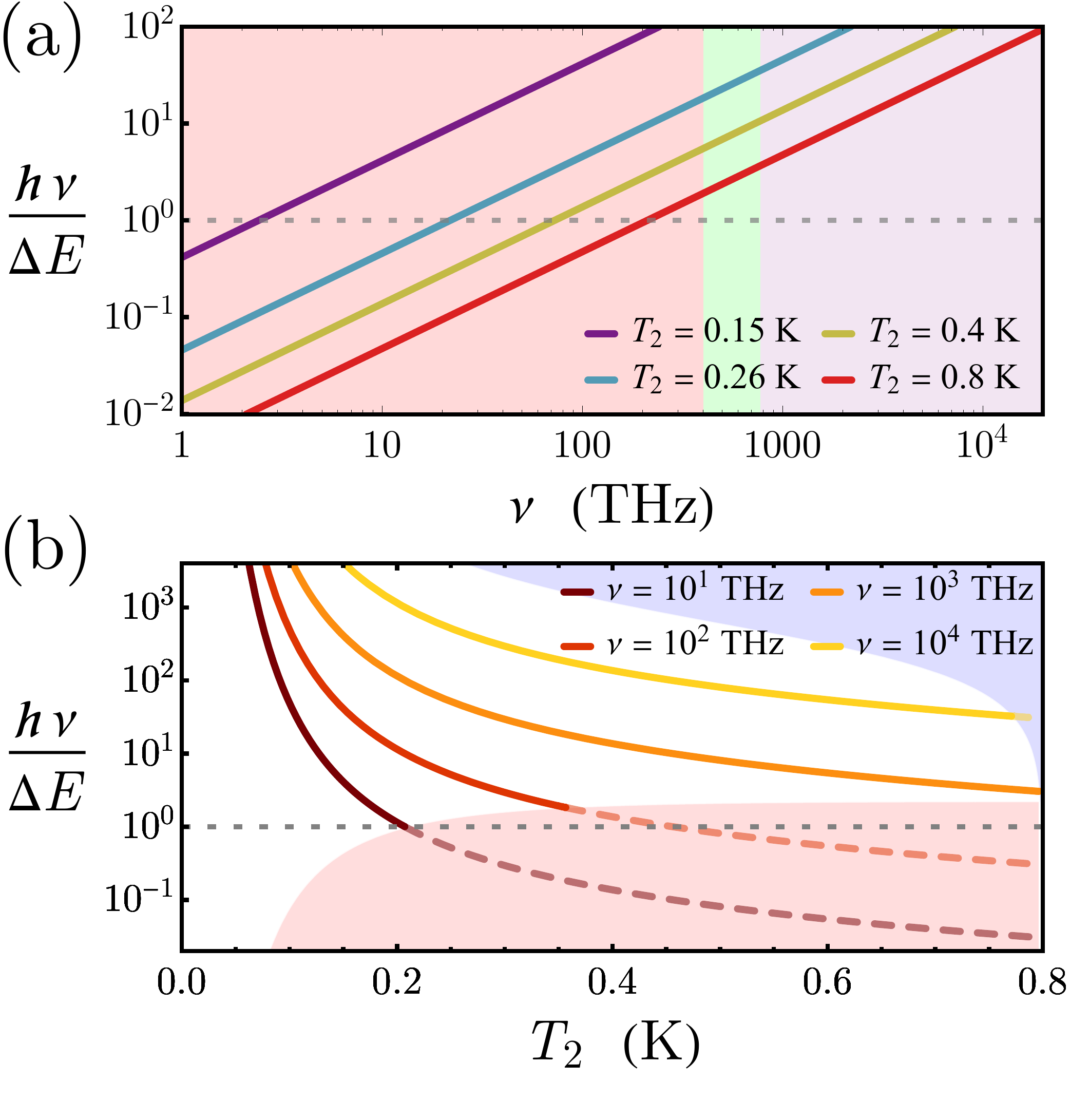}
\caption{(a) Resolving power as a function of the photon frequency at a few temperature. The shaded regions indicate the frequency ranges corresponding to IR (red), visible (green), and UV (purple) light spectrum. (b) Resolving power as a function of the temperatures at a few values of the photon frequency. We are considering an electrode with volume $V_2=1\;\mu\text{m}^3$. The shaded areas indicate forbidden regions, since corresponding to frequencies below $\nu_{\text{min}}$ (light red area) and above $\nu_{\text{max}}$ (light blue area). In both panels, the dashed horizontal line marks the unitary value of the resolving power.}
\label{Fig08}
\end{figure}

To estimate the performance of a calorimeter a relevant figure of merit is the resolving power, which is calculated in the idle state and represents the energy sensitivity for the case of low-energy photons absorption. The resolving power reads~\cite{Vou10,Vir18}
\begin{equation}
\frac{h\nu}{\Delta E}=\frac{h\nu}{4\sqrt{2 \ln 2}\sqrt{k_BT_2^2C_2(T_2)}},
\end{equation}
where $T_2$ is the steady temperature of the absorber and $\Delta E$ is the intrinsic energy resolution of full width at half maximum for a calorimeter, determined by Jonhson-Nyquist thermodynamics fluctuations in energy~\cite{Gia06,Vou10}.
Fig.~\ref{Fig08}(a) shows the resolving power as a function of the photon frequency, $\nu$, at a few temperatures of the electrode $S_2$ with volume $V_2=1\;\mu\text{m}^3$. The horizontal dashed line indicates unitary resolving power. We observe that the resolving power increases linearly with the photon frequency. At the same time, by rising the temperature $T_2$, the increase of the heat capacity $C_2(T_2)$ will reduce the resolving power. We note that at $0.15\;\text{K}$ a resolving power exceeding one results in almost the whole range of frequencies shown in Fig.~\ref{Fig08}(a) (infrared to UV light spectrum). The range of frequencies giving $h\nu/\Delta E>1$ decreases by increasing the temperature. At $T_2=0.26\;\text{K}$ (cyan line), that is roughly the working temperature previously discussed, we obtain $h\nu/\Delta E>1$ only at frequencies above $\sim21\;\text{THz}$. At frequencies above this value, a detector, residing at this temperature and with the chosen detection volume $V_2$, could efficiently work as a calorimeter. Then, in the whole detection frequency range discussed in the previous section we achieve a resolving power larger than 1.

The temperature dependence of the resolving power at a few values of the photon frequency is displayed in Fig.~\ref{Fig08}(b). The shaded areas in this figure indicate forbidden regions, since corresponding to frequencies below $\nu_{\text{min}}$ (light red area) and above $\nu_{\text{max}}$ (light blue area). We note that at a fixed $\nu$ the resolving power monotonically reduces by increasing the temperature, and that the higher $\nu$, the larger the range of temperatures giving $h\nu/\Delta E>1$.

Finally, we note that the preliminary evaluations developed in these sections are performed by omitting both the phononic and the $\text{S}_2\text{I}\text{N}$ thermal relaxation channels, that will be instead considered in the next section, where we deal with the full thermal dynamics of the device. In principle, these terms could modify the dynamical ranges of the calorimeter, but as we will see the estimate given before are still valid.

\subsection{Temperature time evolution}
\label{Sec02d}\vskip-0.2cm

In this section we study the time evolution of the electronic temperature of the electrode $S_2$, when a photon is absorbed.
The results presented hereafter are obtained assuming that the system is always in quasi-equilibrium. This allows us to clearly identify the electronic temperatures of $S_2$ and, at the same time, to describe the time evolution of $T_2$, which is governed by the equation
\begin{eqnarray}
\label{DynamicBalanceEqs}
P_{\text{S}_1\to\text{S}_2}(T_1,T_2,\varphi)&+&P_{\text{rad}}-P_{\text{e-ph,2}}(T_2,T_{\text{bath}})+\\\nonumber
&-&P_{\text{S}_2\to\text{N}}(T_2,T_{\text{bath}})=C_2(T_2)\frac{\mathrm{d} T_2}{\mathrm{d} t}.
\end{eqnarray}
This equation includes all the incoming and outgoing thermal powers in $S_2$. 
We are confident that the predictions of previous equation well represent the detector response on a timescale longer than the intrinsic energy equilibration time of the superconductor.
Indeed, in order to define a proper unique temperature for the floating lead through Eq.~\eqref{DynamicBalanceEqs}, one needs to rely on a clear separation in timescales for the thermalization processes in the system. 
The time the electronic system needs to reach rapidly a thermal distribution can be conservatively estimated as the characteristic \textit{quasiparticle relaxation time}~\cite{Kap76}, $\tau_{\epsilon}$.
This time may be interpreted as a minimum timescale over which it makes sense to speak of an electronic temperature of the absorber and implicitly represents, in a quasi-equilibrium analysis, the minimal time response. 
At the operating temperature $T^0_2=0.258\;\text{K}$, we can estimate for an absorber made, for instance, by Ta, In, or Pb  the quasiparticle relaxation times $\tau_\epsilon^{\text{Ta}}\sim12\;\text{ns}$, $\tau_\epsilon^{\text{In}}\sim5\;\text{ns}$, and $\tau_\epsilon^{\text{Pb}}\sim1.3\;\text{ns}$, respectively. 
So, our detector can be well described by a quasi-equilibrium theory if we investigate it on a timescale longer than the relaxation time of the superconductor.

The dynamical timescale described by Eq.~\eqref{DynamicBalanceEqs} can be estimated in the linear regime, through the \textit{thermal response time}, $\tau_{\text{th}}$, of the device. It can be written as $\tau_{\text{th}}=C_2/(\mathcal{G}+\mathcal{K}_{\text{S}_2\text{I}\text{N}}-\mathcal{K}_{\text{S}_1\text{I}\text{S}_2})$, where $\mathcal{G}$ and $\mathcal{K}_j$ indicate the electron-phonon and electron thermal conductances of the JJ, respectively. The full expressions of the thermal conductances are shown in Appendix~\ref{AppA}. In our case, at the idle temperature $T^0_2=0.258\;\text{K}$ and for the device parameters used previously, one obtains $\tau_{\text{th}}\sim26\;\text{ns}$. We note that $\tau_{\text{th}}$ underestimates the effective time the system takes to restore the initial temperature after a photon absorption. In fact, when we will discuss the time response of the system to two distinct photonic events, we will observe that the system returns back to the idle state in approximatively $0.1-0.2	\;\mu\text{s}$, that is in a time definitively longer than the $\tau_\epsilon$ values estimated before.

Interestingly, we observe that at low values of the resistance of the junction between $\text{S}_2$ and $\text{N}$, the thermalization process is mainly ruled by this thermal relaxation channel, so that $\tau_{\text{th}}$ strongly depends on the resistance $\Rsin$, since $\mathcal{K}_{\text{S}_2\text{I}\text{N}}\propto\Rsin^{-1}$. So, in principle, our detector can potentially react in a very short time depending on the value of $\Rsin$. Alternatively, we could adjust the value of this resistance, in order to make longer the thermal response time, so to better satisfy the required timescale separation.

We can also calculate the typical length over which a system can be considered homogeneous with a thermal electronic distribution, that is the \textit{inelastic scattering length}, $\ell_{\text{in}}=\sqrt{D\tau_\epsilon}$, where $D=\sigma_N/(e^2N_F)$ is the diffusion constant and $\sigma_N$ is the electrical conductivity in the normal state. From the values of $\tau_\epsilon$ estimated previously, we obtain $\ell_{\text{in}}^{\text{Ta}}\sim6\;\mu\text{m}$, $\ell_{\text{in}}^{\text{In}}\sim5\;\mu\text{m}$, and $\ell_{\text{in}}^{\text{Pb}}\sim1.6\;\mu\text{m}$. 
Since in our case this characteristic length is larger the lateral dimension of the floating lead ($\sim1\;\mu\text{m}$) we can consider $S_2$ as thermally homogeneous.

Given these estimations, since the quantity $\tau_\epsilon^{-1}$ represents a sort of minimal bandwidth, we would expect that the analysis is valid for an electrical bandwidth between $0.1-1\;\text{GHz}$, which well fits the requirements of the proposed detection methods based on the kinetic inductance readout (see later).

We model the photon-induced energy diffusion in the superconductor by a Gaussian envelope centered in $t_0$ with standard deviation (in time) $\sigma$, reading as follow
\begin{equation}\label{PhotonPower}
P_{\text{rad}}=\frac{h\nu}{\sqrt{2\pi}\sigma}\exp \left [ -\frac{\left ( t-t_0 \right )^2}{2\sigma^2} \right ].
\end{equation}
We fix the width $\sigma=2\;\text{ns}$ of the Gaussian envelope of the photonic energy diffusion. Since $\sigma\ll\tau_{\text{th}}$, this choice permits to well mimic the short-time behavior of the detector, making also the best from our quasi-equilibrium approach. At the same time, if $\sigma$ is ``too large'', thermal losses can play a role even during the initial thermalization process, making the temperature initially reached by $S_2$ lower than the estimate one obtained previously through Eq.~\eqref{TpeakRelation}.

\begin{figure}[t!!]
\centering
\includegraphics[width=\columnwidth]{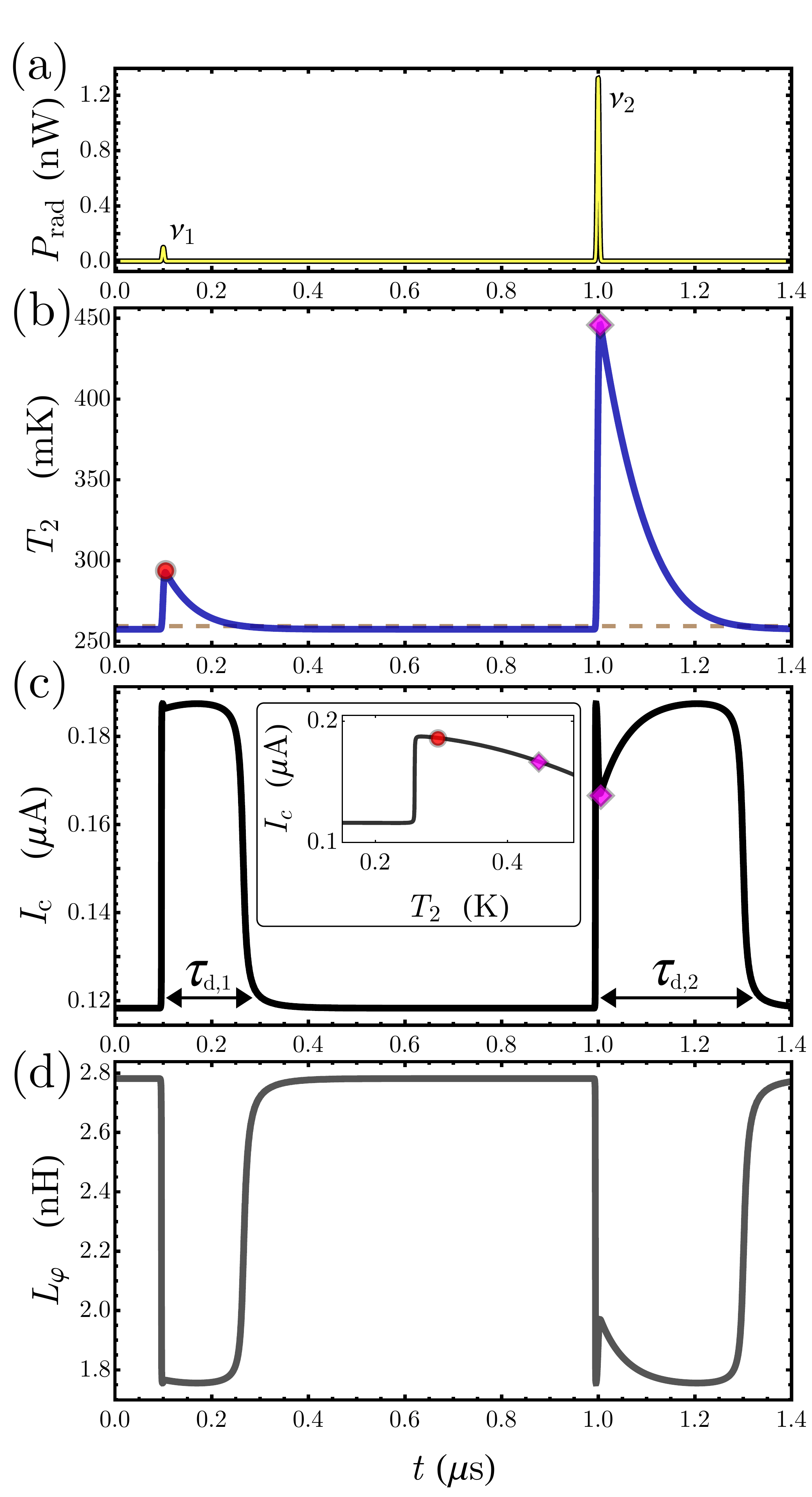}
\caption{(a) Power transmitted by two photons, with $\nu_1=750\;\text{THz}$ and $\nu_2=10^4\;\text{THz}$, absorbed in $S_2$ after $t_0\sim0.1\;\mu\text{s}$ and $t_0\sim1\;\mu\text{s}$, respectively. Variations of temperature (b), critical current (c), and Josephson inductance (d) due to the photonic events described in panel (a). In panel (b), the dashed line marks the temperature $T_2^J$ at which an $I_c$ jump occurs, whereas a red circle and a pink diamond identify the maximum temperatures reached as a result of photonic events. In panel (c), the dead times, $\tau_{\text{d}}$, are also indicated. In the inset: critical current versus $T_2$, where the $I_c$ values at the temperatures marked in panel (b) are also highlighted. The values of other parameters are: $T_{c_1}=1.2\;\text{K}$, $T_{c_2}=0.8\;\text{K}$, $R=1\;\text{k}\Omega$, $V_2=1\;\mu\text{m}^3$, $T_{\text{bath}}=10\;\text{mK}$, and $\varphi=0$.}
\label{Fig09}
\end{figure}

Now, we discuss the response of the detector to a photonic event, where the power of radiation is described by a Gaussian envelope as in Eq.~\eqref{PhotonPower}. In Fig.~\ref{Fig09} we show the behavior of the system when a photon with frequencies $\nu_1=750\;\text{THz}$ (violet light) is absorbed by the detector at the time $t_0\sim0.1\;\mu\text{s}$ [see Fig.~\ref{Fig09}(a)]. After an absorption, during a initial rise the temperature $T_2$ increases reaching a maximum, $T_{2,\text{max}}$, and the critical current undergoes a first jump. Then, due to the thermal contact with the phonon bath and the cooling finger, the electrode $S_2$ recovers its initial idle temperature [see Fig.~\ref{Fig09}(b)]. During the thermal evolution following a photon absorption, the condition $T_2=T_2^J$ is satisfied twice, when the temperature is increasing and then when it is decreasing [see dashed line in Fig.~\ref{Fig09}(b)]. This means that a single photonic event causes two subsequent $I_c$ jumps of opposite sign. 
In fact, while the temperature reduces towards the idle value, the critical current goes through a second jump when $T_2=T_2^J$. 

The distance in time between the two subsequent $I_c$ jumps induced by a photonic event can be used to define the dead time, $\tau_{\text{d}}$, of the device [see Fig.~\ref{Fig09}(c)]. This is the time frame in which the detector cannot be used to reveal the arrival of a following incident photon. In fact, once a transition induced by a photon with enough energy has occurred, a further photon-induced temperature enhancement would not trigger another $I_c$ jump, unless the system has already switched back to its idle state~\footnote{More specifically, the detector cannot reveal the arrival timing of a second photon, but it will anyway catch, due to the longer dead time, the additional energy associated to it.}. 

In Fig.~\ref{Fig09} we show also the response of the device due a second photonic event with frequency $\nu_2=10^4\;\text{THz}$ (extreme UV), absorbed at $t_0\sim1\;\mu\text{s}$, that is when the system relaxed to its idle state after the first photonic event. We observe that the higher the photon energy, the higher the maximum temperature, $T_{2,\text{max}}$, reached by $S_2$. During the initial temperature rise, after the increase of $I_c$, we note a following dip, corresponding to $I_c(T_{2,\text{max}})$. 
In fact, during this evolution, the temperature $T_2$ increases reaching its maximum, and $I_c$ first suddenly increases and then it rapidly reduces, reaching the value $I_c(T_{2,\text{max}})$ [see Fig.~\ref{Fig09}(c)]. Notably, this value of the critical current is lower than its maximum value, see the current profile shown in the inset of Fig.~\ref{Fig09}(c), and this is the reason for the appearance of the tight peak in the $I_c$ response once the photon has been absorbed. Both $T_{2,\text{max}}$ and $I_c(T_{2,\text{max}})$ following the absorption of the photons with frequencies $\nu_1$ and $\nu_2$, are marked in Figs.~\ref{Fig09}(b) and~(c) by a red circle and a pink diamond, respectively. 
When the temperature reduces from $T_{2,\text{max}}$ towards the idle state, one can imagine to ``shift back'' the pink dot along the $I_c$ profile shown in the inset. Then, during this evolution, $I_c$ first increases again, reaching the maximum, and it jumps back to the idle value.

\begin{figure}[t!!]
\centering
\includegraphics[width=\columnwidth]{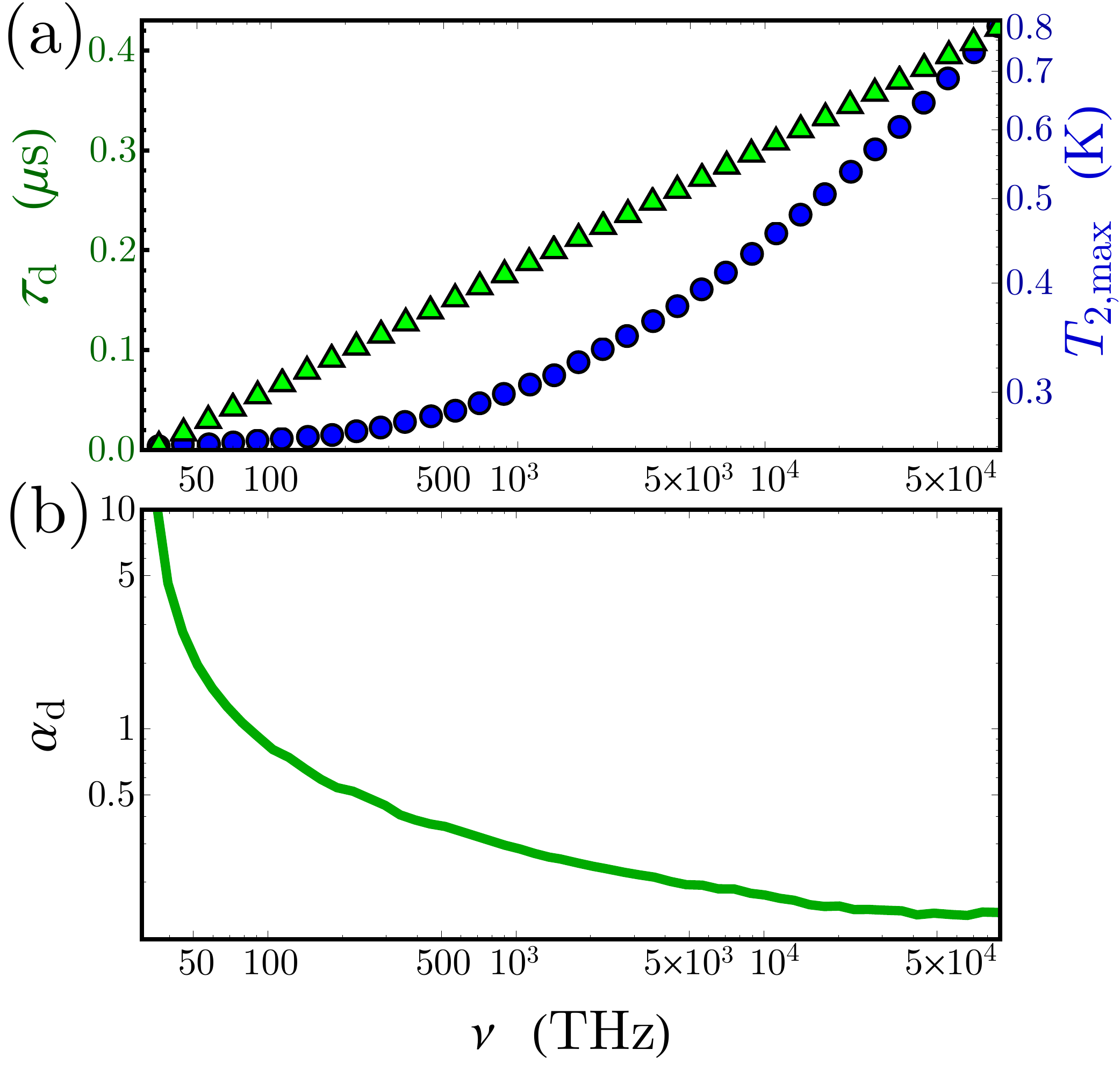}
\caption{(a), Dead time, $\tau_{\text{d}}$, (left axis, green triangles) and maximum temperature, $T_{2,\text{max}}$, (right axis, blue circles) as a function of the photon frequency $\nu$. (b), adimensional figure of merit of the calorimeter $\alpha_{\text{d}}=\frac{\nu}{\tau_{\text{d}}}\frac{\mathrm{d}\tau_{\text{d}} }{\mathrm{d} \nu}$ as a function of $\nu$. The values of other parameters are the same used in Fig.~\ref{Fig09}.}
\label{Fig10}
\end{figure}

Interestingly, we observe also that the higher the photon energy, the higher $T_{2,\text{max}}$ and therefore the longer the time that the system takes to approach its idle state, that is the longer the dead time $\tau_{\text{d}}$ [see Figs.~\ref{Fig09}(c)].
Since $T_{2,\text{max}}$ depends on $\nu$, we can use the dead time as a probe to find the frequency of the absorbed photons. In Fig.~\ref{Fig10}(a) we show the behaviour of both the dead time, $\tau_{\text{d}}$, (left axis, green triangles) and the maximum temperature, $T_{2,\text{max}}$, (right axis, blue circles) as a function of the photon frequency $\nu$. We note that both quantities grow monotonically by increasing the photon frequency. Interestingly, the dead time $\tau_{\text{d}}$ increases exponentially with the frequency, according to the exponential decay of the temperature after the absorption. 
At the same time, $T_{2,\text{max}}$ follows exactly the prediction as calculated through Eq.~\ref{TpeakRelation} and shown in Fig.~\ref{Fig06}.
We observe also that there is no mismatch between the detection frequency range computed in this case and that one discussed in Sec.~\ref{Sec02aa}, although we are now taking into account also the thermal relaxation channels. This is because we are assuming a much shorter photon-induced energy relaxation time with respect to the thermal response time.

Finally, according to the nonlinear behavior of $\tau_{\text{d}}$, we can define another adimensional figure of merit of our calorimeter, namely, the logarithmic derivative $\alpha_{\text{d}}=\frac{\nu}{\tau_{\text{d}}}\frac{\mathrm{d}\tau_{\text{d}} }{\mathrm{d} \nu}$. 
This quantity is larger at the low frequencies, but it tends to drastically reduces in the UV part of the frequency spectrum. This means that the capability of the device to discern the photon frequency by measuring the dead time is higher in the low part of the detection frequency range [see Fig.~\ref{Fig10}(b)].

From the dead time shown in Fig.~\ref{Fig10}(a), one can derive the detection rate of photons for the device. In the case of wide spectrum detection, with the parameters considered in our work, the maximal detection rate is $\tau^{-1}_{\text{d}}(\nu_{\text{max}})\simeq2.5\;\text{MHz}$, whereas in the near infrared band we obtain $\tau^{-1}_{\text{d}}(430\;\text{THz})\simeq7\;\text{MHz}$.

Inasmuch as the dead time is proportional to the exponential of the energy absorbed, in the case of monochromatic radiation our device can show unique photon-number-resolving detection capabilities.

Finally, we observe that one could in principle include thermodynamic fluctuations, estimated through Eq.~\eqref{TemperatureFluctuation}, into the thermal model Eq.~\eqref{DynamicBalanceEqs} (e.g., Ref.~\cite{Bra18}). This improvement of the numerical approach will eventually result in a noisy thermal evolution (and therefore in a noisy evolution of both the critical current and the kinetic inductance), but we not expect relevant noise-triggered effects since we conservatively set a large enough detection threshold keeping minimal dark counts.

\subsection{The readout}
\label{Sec02e}\vskip-0.2cm

\begin{figure}[t!!]
\centering
\includegraphics[width=\columnwidth]{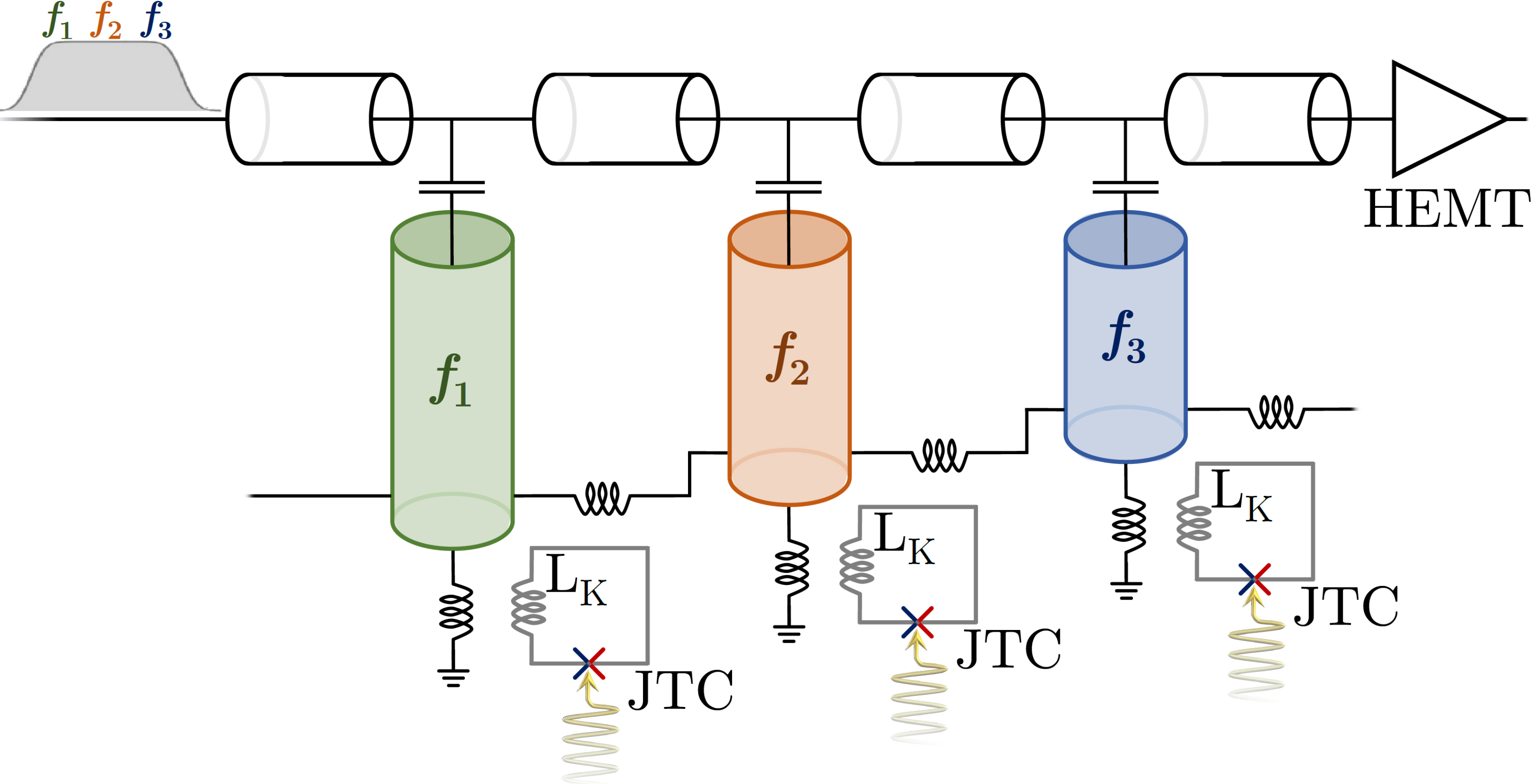}
\caption{Schematic representation of a multiplexing scheme for JTCs, along the lines of the SQUID-based multiplexing scheme implemented in Ref.~\cite{Mat17}.}
\label{Fig11}
\end{figure}

Detection of photonic events can be done by reading the Josephson kinetic inductance, $L_{\varphi}$, of the junction. In fact, the behavior of $I_c$ reflects on the kinetic inductance according to~\cite{Bar82,Lik86} 
\begin{equation}\label{JosephsonInductance}
L_{\varphi}=\frac{\Phi_0}{2\pi}\left (\frac{\partial I_{\varphi}}{\partial \varphi} \right )^{-1}=\frac{\Phi_0}{2\pi}\frac{1}{\cos\varphi\,I_c},
\end{equation}
as shown in Fig.~\ref{Fig09}(d). Here, $I_\varphi$ is the typical current-phase relation of a tunnel JJ.
Thus, the readout of the Josephson kinetic inductance can be performed dispersively through an LC resonator inductively coupled to the JJ~\cite{Gov14,Gov16}. In this readout scheme the modification of the inductance can be measured through a shift of the circuit's transmission or reflection resonance~\cite{Day03}. Notably, multiplex readout is permitted, along the lines of the SQUID multiplexing scheme discussed in Refs.~\cite{Ull15,Mat17}. In this scheme, a rf-SQUID is inductively coupled to a resonator, with a specific characteristic frequency. The change in the kinetic inductance of the SQUID affects the frequency response of each resonator. Then, all the resonators are coupled to a common feedline towards a high-bandwidth cryogenic HEMT amplifier~\cite{Mat17}. Each resonator is tuned at a different frequency, so that all the sensors can be read simultaneously by applying a wide spectrum of probing frequencies to the feedline. Unlike the scheme proposed in Refs.~\cite{Ull15,Mat17}, in the present case the rf-SQUID is exactly our radiation sensor, where the SQUID ring is formed by the temperature-biased asymmetric JJ used for the photon detection and the superconducting ring utilized for the phase-biasing [see Fig.~\ref{Fig01}]. A schematic representation showing just three multiplexing channels is depicted in Fig.~\ref{Fig11}. In this figure, we show also a modulation flux bias line used to tune the SQUID fluxes. When the ring inductance of each detector is made negligible, the kinetic inductance, $L_{\text{K}}$, of the SQUID is dominated by the Josephson inductance $L_{\varphi}$. Nevertheless, the investigation of the optimal multiplexing strategy is a topic for further researches.

\section{Conclusions}
\label{Sec03}\vskip-0.2cm

In summary, we propose a threshold calorimeter based on the peculiar behavior of the critical current, $I_c$, of a temperature-biased tunnel Josephson junction made by different superconductors. In fact, the step-like variation with the temperature of $I_c$, already shown in Ref.~\cite{GuaBra18}, can allow us to design a single-photon threshold detector in which the sensing element is one lead of the junction. Then, the absorption of a photon produces an enhancement of the electronic temperature, which can induce a measurable sudden increment of the critical current.

The conceived device is inherently energy resolving, and can be also engineered to determine the photon number in the case of a monochromatic source of light. In order to prevent an unreliable absorber temperature readout and minimize dark counts, we investigated the thermodynamics fluctuations in the superconductor. This analysis allows us to settle the idle temperatures of the device. Then, we discussed the essential figure of merit of this type of detector, i.e., the resolving power, in order to find the optimal detection design. With our choice of realistic parameters for the setup investigated, the proposed sensor can efficiently detect photons of frequencies $\nu\in[30-9\times10^4]\;\text{THz}$. Our sensor is also able to determine the photon frequency by measuring the dead time of the detector. 
Interestingly, we observe that both the capability to discriminate the photon frequency through the dead time and the detection rate of the sensor are stronger at low frequencies.

We propose a non-invasive readout scheme based on the modification of the Josephson kinetic inductance caused by a photon-induced temperature enhancement. This quantity can be dispersively read via a LC resonator inductively coupled to the detector. This readout scheme is suitable for a multiplexing configuration.

We note that our proposal evinces some similarities with other single-photon detectors based on the photon-induced $I_c$ changes in a proximized nanowire~\cite{Vou10,Gov16,Vir18,Sol18}. Conversely, in our detection scheme the absorbing element is an electrode of an asymmetric JJ and the phenomenon exploited is the jump in $I_c$ by changing the temperature. Therefore, the strength of our proposal resides in a strong sensitivity due to the steep response of $I_c$ to a photonic event. Moreover, the detection is not performed at extremely low temperature, resulting both in a fast thermal response and a shorter dead time of the detector. Markedly, our detector represents an interesting combination between different types of superconducting single-photon and calorimetric devices. It has the potential sensitivity of superconducting tunnel junction detectors, however being not affected by Johnson-Nyquist noise, since working in a dissipationless regime. Moreover, the proposed detector has potentially the energy sensitivity of proximity-based detectors, with a reduced dead time, due to higher operating temperatures. 

Finally, we observe that the characteristic timescales of the system depend on the choice of the superconductor used as absorbing element and on the characteristics of the metallic cooling finger. Moreover, the use of superconductors with higher $T_c$’s would permit higher working temperatures, resulting in a further reduction of both the thermal response~\cite{Gua17} and the quasiparticle relaxation~\cite{Kap76,GuaSolBra18} times. So, a careful material selection could outperform the conservative estimate adopted in the above design study.

\begin{acknowledgments}

We acknowledge F. Paolucci for useful discussions. C.G., A.B., and F.G. acknowledge the European Research Council under the European Union's Seventh Framework Program (FP7/2007-2013)/ERC Grant agreement No.~615187-COMANCHE and the Tuscany Region under the PAR FAS 2007-2013, FAR-FAS 2014 call, project SCIADRO, for financial support.
P.S. and A.B. have received funding from the European Union FP7/2007-2013 under REA Grant agreement No. 630925 -- COHEAT. 
A.B. acknowledges the CNR-CONICET cooperation programme ``Energy conversion in quantum nanoscale hybrid devices'' and the Royal Society though the International Exchanges between the UK and Italy (grant IES R3 170054).
\end{acknowledgments}

\appendix

\section{Timescales of the thermalization process.}
\label{AppA}

The thermal response time $\tau_{\text{th}}$ can be estimated by first-order expanding the heat current terms in Eq.~\eqref{Pt} in the idle state of the system~\cite{GuaSol18}. Specifically, it reads $\tau_{\text{th}}=C_2/(\mathcal{G}+\mathcal{K}_{\text{S}_2\text{I}\text{N}}-\mathcal{K}_{\text{S}_1\text{I}\text{S}_2})$, where $\mathcal{G}$ and $\mathcal{K}$ indicate the electron-phonon and electron thermal conductances of the JJ, respectively. 
These conductances can be obtained through the first derivatives of the heat power densities in Eqs.~\eqref{StatBalanceEqs} and~\eqref{DynamicBalanceEqs}, calculated at a steady electronic temperature $T_e$. Specifically the electron-phonon thermal conductance reads~\cite{Vir18}
\begin{eqnarray}
&\mathcal{G}_2(T_e)=\frac{\partial P_{e\text{-ph},2}}{\partial T_e} =\qquad\qquad\qquad\qquad\qquad\qquad\\\nonumber
&=\frac{5\Sigma_2 V_2}{960\zeta (5)k_B^6T_e^6}\displaystyle\iint_{-\infty}^{\infty}\frac{dEd\varepsilon E\left | \varepsilon \right |^3M^2_{E,E-\varepsilon }}{\sinh\frac{\varepsilon }{2k_BT_e}\cosh\frac{E }{2k_BT_e}\cosh\frac{E-\varepsilon }{2k_BT_e}},
\label{ephconductance}
\end{eqnarray}
while the electron thermal conductances read~\cite{Mar14}
\begin{eqnarray}
&\mathcal{K}_{\text{S}_1\text{I}\text{S}_2}(T_e)=\frac{\partial P_{\text{S}_1\to\text{S}_2}}{\partial T_e} =\frac{1}{2e^2k_BT_e^2R}\displaystyle\int_{0}^{\infty}\frac{d\varepsilon \varepsilon^2}{\cosh^2\frac{\varepsilon }{2k_BT_e}}\times\qquad\\\nonumber
&\times\Big [\mathcal{N}_1(\varepsilon,T_e)\mathcal{N}_2(\varepsilon,T_e)-\mathcal{M}_1(\varepsilon,T_e)\mathcal{M}_2(\varepsilon,T_e)\cos\varphi\Big ],
\label{econductance_SIS}
\end{eqnarray}
and
\begin{eqnarray}
\mathcal{K}_{\text{S}_2\text{I}\text{N}}(T_e)&=&\frac{\partial P_{\text{S}_2\to\text{N}}}{\partial T_e} =\frac{1}{2e^2k_BT_e^2\Rsin}\times\\\nonumber
&\times&\displaystyle\int_{0}^{\infty}\frac{d\varepsilon \varepsilon^2}{\cosh^2\frac{\varepsilon }{2k_BT_e}}\mathcal{N}_2(\varepsilon,T_e)
\label{econductance_SIN}
\end{eqnarray}
where $\mathcal{N}_j\left ( \varepsilon ,T \right )=\left | \text{Re}\left [ \frac{ \varepsilon +i\Gamma_j}{\sqrt{(\varepsilon +i\Gamma_j) ^2-\Delta _j\left ( T \right )^2}} \right ] \right |$ and $\mathcal{M}_j\left ( \varepsilon ,T \right )=\left |\text{Im}\left [\frac{ - i\Delta _j\left ( T \right )}{\sqrt{\left (\varepsilon+ i\Gamma_j \right ) ^2-\Delta _j\left ( T \right )^2}} \right ] \right |$.


\begin{thebibliography}{69}%
\makeatletter
\providecommand \@ifxundefined [1]{%
 \@ifx{#1\undefined}
}%
\providecommand \@ifnum [1]{%
 \ifnum #1\expandafter \@firstoftwo
 \else \expandafter \@secondoftwo
 \fi
}%
\providecommand \@ifx [1]{%
 \ifx #1\expandafter \@firstoftwo
 \else \expandafter \@secondoftwo
 \fi
}%
\providecommand \natexlab [1]{#1}%
\providecommand \enquote  [1]{``#1''}%
\providecommand \bibnamefont  [1]{#1}%
\providecommand \bibfnamefont [1]{#1}%
\providecommand \citenamefont [1]{#1}%
\providecommand \href@noop [0]{\@secondoftwo}%
\providecommand \href [0]{\begingroup \@sanitize@url \@href}%
\providecommand \@href[1]{\@@startlink{#1}\@@href}%
\providecommand \@@href[1]{\endgroup#1\@@endlink}%
\providecommand \@sanitize@url [0]{\catcode `\\12\catcode `\$12\catcode
  `\&12\catcode `\#12\catcode `\^12\catcode `\_12\catcode `\%12\relax}%
\providecommand \@@startlink[1]{}%
\providecommand \@@endlink[0]{}%
\providecommand \url  [0]{\begingroup\@sanitize@url \@url }%
\providecommand \@url [1]{\endgroup\@href {#1}{\urlprefix }}%
\providecommand \urlprefix  [0]{URL }%
\providecommand \Eprint [0]{\href }%
\providecommand \doibase [0]{http://dx.doi.org/}%
\providecommand \selectlanguage [0]{\@gobble}%
\providecommand \bibinfo  [0]{\@secondoftwo}%
\providecommand \bibfield  [0]{\@secondoftwo}%
\providecommand \translation [1]{[#1]}%
\providecommand \BibitemOpen [0]{}%
\providecommand \bibitemStop [0]{}%
\providecommand \bibitemNoStop [0]{.\EOS\space}%
\providecommand \EOS [0]{\spacefactor3000\relax}%
\providecommand \BibitemShut  [1]{\csname bibitem#1\endcsname}%
\let\auto@bib@innerbib\@empty
\bibitem [{\citenamefont {Giazotto}\ \emph {et~al.}(2006)\citenamefont
  {Giazotto}, \citenamefont {Heikkil\"a}, \citenamefont {Luukanen},
  \citenamefont {Savin},\ and\ \citenamefont {Pekola}}]{Gia06}%
  \BibitemOpen
  \bibfield  {author} {\bibinfo {author} {\bibfnamefont {F.}~\bibnamefont
  {Giazotto}}, \bibinfo {author} {\bibfnamefont {T.~T.}\ \bibnamefont
  {Heikkil\"a}}, \bibinfo {author} {\bibfnamefont {A.}~\bibnamefont
  {Luukanen}}, \bibinfo {author} {\bibfnamefont {A.~M.}\ \bibnamefont {Savin}},
  \ and\ \bibinfo {author} {\bibfnamefont {J.~P.}\ \bibnamefont {Pekola}},\
  }\bibfield  {title} {\enquote {\bibinfo {title} {Opportunities for
  mesoscopics in thermometry and refrigeration: Physics and applications},}\
  }\href {\doibase 10.1103/RevModPhys.78.217} {\bibfield  {journal} {\bibinfo
  {journal} {Rev. Mod. Phys.}\ }\textbf {\bibinfo {volume} {78}},\ \bibinfo
  {pages} {217--274} (\bibinfo {year} {2006})}\BibitemShut {NoStop}%
\bibitem [{\citenamefont {Ullom}\ and\ \citenamefont {Bennett}(2015)}]{Ull15}%
  \BibitemOpen
  \bibfield  {author} {\bibinfo {author} {\bibfnamefont {J.~N.}\ \bibnamefont
  {Ullom}}\ and\ \bibinfo {author} {\bibfnamefont {D.~A.}\ \bibnamefont
  {Bennett}},\ }\bibfield  {title} {\enquote {\bibinfo {title} {Review of
  superconducting transition-edge sensors for x--ray and gamma--ray
  spectroscopy},}\ }\href {http://stacks.iop.org/0953-2048/28/i=8/a=084003}
  {\bibfield  {journal} {\bibinfo  {journal} {Supercond. Sci. Technol.}\
  }\textbf {\bibinfo {volume} {28}},\ \bibinfo {pages} {084003} (\bibinfo
  {year} {2015})}\BibitemShut {NoStop}%
\bibitem [{\citenamefont {Dauler}\ \emph {et~al.}(2014)\citenamefont {Dauler},
  \citenamefont {Matthew}, \citenamefont {Kerman}, \citenamefont {Marsili},
  \citenamefont {Miki}, \citenamefont {Nam}, \citenamefont {Shaw},
  \citenamefont {Terai}, \citenamefont {Verma},\ and\ \citenamefont
  {Yamashita}}]{Dau14}%
  \BibitemOpen
  \bibfield  {author} {\bibinfo {author} {\bibfnamefont {E.~A.}\ \bibnamefont
  {Dauler}}, \bibinfo {author} {\bibfnamefont {E.~G.}\ \bibnamefont {Matthew}},
  \bibinfo {author} {\bibfnamefont {A.~J.}\ \bibnamefont {Kerman}}, \bibinfo
  {author} {\bibfnamefont {F.}~\bibnamefont {Marsili}}, \bibinfo {author}
  {\bibfnamefont {S.}~\bibnamefont {Miki}}, \bibinfo {author} {\bibfnamefont
  {S.~W.}\ \bibnamefont {Nam}}, \bibinfo {author} {\bibfnamefont {M.~D.}\
  \bibnamefont {Shaw}}, \bibinfo {author} {\bibfnamefont {H.}~\bibnamefont
  {Terai}}, \bibinfo {author} {\bibfnamefont {V.~B.}\ \bibnamefont {Verma}}, \
  and\ \bibinfo {author} {\bibfnamefont {T.}~\bibnamefont {Yamashita}},\
  }\bibfield  {title} {\enquote {\bibinfo {title} {Review of superconducting
  nanowire single-photon detector system design options and demonstrated
  performance},}\ }\href {\doibase 10.1117/1.OE.53.8.081907} {\bibfield
  {journal} {\bibinfo  {journal} {Optical Engineering}\ }\textbf {\bibinfo
  {volume} {53}},\ \bibinfo {pages} {081907} (\bibinfo {year}
  {2014})}\BibitemShut {NoStop}%
\bibitem [{\citenamefont {Pekola}\ \emph {et~al.}(2013)\citenamefont {Pekola},
  \citenamefont {Solinas}, \citenamefont {Shnirman},\ and\ \citenamefont
  {Averin}}]{Pek13}%
  \BibitemOpen
  \bibfield  {author} {\bibinfo {author} {\bibfnamefont {J.~P.}\ \bibnamefont
  {Pekola}}, \bibinfo {author} {\bibfnamefont {P.}~\bibnamefont {Solinas}},
  \bibinfo {author} {\bibfnamefont {A.}~\bibnamefont {Shnirman}}, \ and\
  \bibinfo {author} {\bibfnamefont {D.~V.}\ \bibnamefont {Averin}},\ }\bibfield
   {title} {\enquote {\bibinfo {title} {Calorimetric measurement of work in a
  quantum system},}\ }\href {http://stacks.iop.org/1367-2630/15/i=11/a=115006}
  {\bibfield  {journal} {\bibinfo  {journal} {New J. Phys.}\ }\textbf {\bibinfo
  {volume} {15}},\ \bibinfo {pages} {115006} (\bibinfo {year}
  {2013})}\BibitemShut {NoStop}%
\bibitem [{\citenamefont {Donvil}\ \emph {et~al.}(2018)\citenamefont {Donvil},
  \citenamefont {Muratore-Ginanneschi}, \citenamefont {Pekola},\ and\
  \citenamefont {Schwieger}}]{Don18}%
  \BibitemOpen
  \bibfield  {author} {\bibinfo {author} {\bibfnamefont {B.}~\bibnamefont
  {Donvil}}, \bibinfo {author} {\bibfnamefont {P.}~\bibnamefont
  {Muratore-Ginanneschi}}, \bibinfo {author} {\bibfnamefont {J.~P.}\
  \bibnamefont {Pekola}}, \ and\ \bibinfo {author} {\bibfnamefont
  {K.}~\bibnamefont {Schwieger}},\ }\bibfield  {title} {\enquote {\bibinfo
  {title} {Model for calorimetric measurements in an open quantum system},}\
  }\href {\doibase 10.1103/PhysRevA.97.052107} {\bibfield  {journal} {\bibinfo
  {journal} {Phys. Rev. A}\ }\textbf {\bibinfo {volume} {97}},\ \bibinfo
  {pages} {052107} (\bibinfo {year} {2018})}\BibitemShut {NoStop}%
\bibitem [{\citenamefont {Brange}\ \emph {et~al.}(2018)\citenamefont {Brange},
  \citenamefont {Samuelsson}, \citenamefont {Karimi},\ and\ \citenamefont
  {Pekola}}]{Bra18}%
  \BibitemOpen
  \bibfield  {author} {\bibinfo {author} {\bibfnamefont {F.}~\bibnamefont
  {Brange}}, \bibinfo {author} {\bibfnamefont {P.}~\bibnamefont {Samuelsson}},
  \bibinfo {author} {\bibfnamefont {B.}~\bibnamefont {Karimi}}, \ and\ \bibinfo
  {author} {\bibfnamefont {J.~P.}\ \bibnamefont {Pekola}},\ }\bibfield  {title}
  {\enquote {\bibinfo {title} {Nanoscale quantum calorimetry with electronic
  temperature fluctuations},}\ }\href {\doibase 10.1103/PhysRevB.98.205414}
  {\bibfield  {journal} {\bibinfo  {journal} {Phys. Rev. B}\ }\textbf {\bibinfo
  {volume} {98}},\ \bibinfo {pages} {205414} (\bibinfo {year}
  {2018})}\BibitemShut {NoStop}%
\bibitem [{\citenamefont {Gasparinetti}\ \emph {et~al.}(2015)\citenamefont
  {Gasparinetti}, \citenamefont {Viisanen}, \citenamefont {Saira},
  \citenamefont {Faivre}, \citenamefont {Arzeo}, \citenamefont {Meschke},\ and\
  \citenamefont {Pekola}}]{Gas15}%
  \BibitemOpen
  \bibfield  {author} {\bibinfo {author} {\bibfnamefont {S.}~\bibnamefont
  {Gasparinetti}}, \bibinfo {author} {\bibfnamefont {K.~L.}\ \bibnamefont
  {Viisanen}}, \bibinfo {author} {\bibfnamefont {O.-P.}\ \bibnamefont {Saira}},
  \bibinfo {author} {\bibfnamefont {T.}~\bibnamefont {Faivre}}, \bibinfo
  {author} {\bibfnamefont {M.}~\bibnamefont {Arzeo}}, \bibinfo {author}
  {\bibfnamefont {M.}~\bibnamefont {Meschke}}, \ and\ \bibinfo {author}
  {\bibfnamefont {J.~P.}\ \bibnamefont {Pekola}},\ }\bibfield  {title}
  {\enquote {\bibinfo {title} {Fast electron thermometry for ultrasensitive
  calorimetric detection},}\ }\href {\doibase 10.1103/PhysRevApplied.3.014007}
  {\bibfield  {journal} {\bibinfo  {journal} {Phys. Rev. Applied}\ }\textbf
  {\bibinfo {volume} {3}},\ \bibinfo {pages} {014007} (\bibinfo {year}
  {2015})}\BibitemShut {NoStop}%
\bibitem [{\citenamefont {Saira}\ \emph {et~al.}(2016)\citenamefont {Saira},
  \citenamefont {Zgirski}, \citenamefont {Viisanen}, \citenamefont {Golubev},\
  and\ \citenamefont {Pekola}}]{Sai16}%
  \BibitemOpen
  \bibfield  {author} {\bibinfo {author} {\bibfnamefont {O.-P.}\ \bibnamefont
  {Saira}}, \bibinfo {author} {\bibfnamefont {M.}~\bibnamefont {Zgirski}},
  \bibinfo {author} {\bibfnamefont {K.~L.}\ \bibnamefont {Viisanen}}, \bibinfo
  {author} {\bibfnamefont {D.~S.}\ \bibnamefont {Golubev}}, \ and\ \bibinfo
  {author} {\bibfnamefont {J.~P.}\ \bibnamefont {Pekola}},\ }\bibfield  {title}
  {\enquote {\bibinfo {title} {Dispersive thermometry with a Josephson junction
  coupled to a resonator},}\ }\href {\doibase 10.1103/PhysRevApplied.6.024005}
  {\bibfield  {journal} {\bibinfo  {journal} {Phys. Rev. Applied}\ }\textbf
  {\bibinfo {volume} {6}},\ \bibinfo {pages} {024005} (\bibinfo {year}
  {2016})}\BibitemShut {NoStop}%
\bibitem [{\citenamefont {Wang}\ \emph {et~al.}(2018)\citenamefont {Wang},
  \citenamefont {Saira},\ and\ \citenamefont {Pekola}}]{Wan18}%
  \BibitemOpen
  \bibfield  {author} {\bibinfo {author} {\bibfnamefont {L.~B.}\ \bibnamefont
  {Wang}}, \bibinfo {author} {\bibfnamefont {O.-P.}\ \bibnamefont {Saira}}, \
  and\ \bibinfo {author} {\bibfnamefont {J.~P.}\ \bibnamefont {Pekola}},\
  }\bibfield  {title} {\enquote {\bibinfo {title} {Fast thermometry with a
  proximity Josephson junction},}\ }\href {\doibase 10.1063/1.5010236}
  {\bibfield  {journal} {\bibinfo  {journal} {Appl. Phys. Lett.}\ }\textbf
  {\bibinfo {volume} {112}},\ \bibinfo {pages} {013105} (\bibinfo {year}
  {2018})}\BibitemShut {NoStop}%
\bibitem [{\citenamefont {Zgirski}\ \emph {et~al.}(2018)\citenamefont
  {Zgirski}, \citenamefont {Foltyn}, \citenamefont {Savin}, \citenamefont
  {Norowski}, \citenamefont {Meschke},\ and\ \citenamefont {Pekola}}]{Zgi18}%
  \BibitemOpen
  \bibfield  {author} {\bibinfo {author} {\bibfnamefont {M.}~\bibnamefont
  {Zgirski}}, \bibinfo {author} {\bibfnamefont {M.}~\bibnamefont {Foltyn}},
  \bibinfo {author} {\bibfnamefont {A.}~\bibnamefont {Savin}}, \bibinfo
  {author} {\bibfnamefont {K.}~\bibnamefont {Norowski}}, \bibinfo {author}
  {\bibfnamefont {M.}~\bibnamefont {Meschke}}, \ and\ \bibinfo {author}
  {\bibfnamefont {J.}~\bibnamefont {Pekola}},\ }\bibfield  {title} {\enquote
  {\bibinfo {title} {Nanosecond thermometry with Josephson junctions},}\ }\href
  {\doibase 10.1103/PhysRevApplied.10.044068} {\bibfield  {journal} {\bibinfo
  {journal} {Phys. Rev. Applied}\ }\textbf {\bibinfo {volume} {10}},\ \bibinfo
  {pages} {044068} (\bibinfo {year} {2018})}\BibitemShut {NoStop}%
\bibitem [{\citenamefont {Karimi}\ and\ \citenamefont {Pekola}(2018)}]{Kar18}%
  \BibitemOpen
  \bibfield  {author} {\bibinfo {author} {\bibfnamefont {B.}~\bibnamefont
  {Karimi}}\ and\ \bibinfo {author} {\bibfnamefont {J.~P.}\ \bibnamefont
  {Pekola}},\ }\bibfield  {title} {\enquote {\bibinfo {title} {Noninvasive
  thermometer based on the zero-bias anomaly of a superconducting junction for
  ultrasensitive calorimetry},}\ }\href {\doibase
  10.1103/PhysRevApplied.10.054048} {\bibfield  {journal} {\bibinfo  {journal}
  {Phys. Rev. Applied}\ }\textbf {\bibinfo {volume} {10}},\ \bibinfo {pages}
  {054048} (\bibinfo {year} {2018})}\BibitemShut {NoStop}%
\bibitem [{\citenamefont {Silaev}\ \emph {et~al.}(2014)\citenamefont {Silaev},
  \citenamefont {Heikkil\"a},\ and\ \citenamefont {Virtanen}}]{Sil14}%
  \BibitemOpen
  \bibfield  {author} {\bibinfo {author} {\bibfnamefont {M.}~\bibnamefont
  {Silaev}}, \bibinfo {author} {\bibfnamefont {T.~T.}\ \bibnamefont
  {Heikkil\"a}}, \ and\ \bibinfo {author} {\bibfnamefont {P.}~\bibnamefont
  {Virtanen}},\ }\bibfield  {title} {\enquote {\bibinfo {title}
  {Lindblad-equation approach for the full counting statistics of work and heat
  in driven quantum systems},}\ }\href {\doibase 10.1103/PhysRevE.90.022103}
  {\bibfield  {journal} {\bibinfo  {journal} {Phys. Rev. E}\ }\textbf {\bibinfo
  {volume} {90}},\ \bibinfo {pages} {022103} (\bibinfo {year}
  {2014})}\BibitemShut {NoStop}%
\bibitem [{\citenamefont {Takesue}\ \emph {et~al.}(2007)\citenamefont
  {Takesue}, \citenamefont {Nam}, \citenamefont {Zhang}, \citenamefont
  {Hadfield}, \citenamefont {Honjo}, \citenamefont {Tamaki},\ and\
  \citenamefont {Yamamoto}}]{Tak07}%
  \BibitemOpen
  \bibfield  {author} {\bibinfo {author} {\bibfnamefont {H.}~\bibnamefont
  {Takesue}}, \bibinfo {author} {\bibfnamefont {S.~W.}\ \bibnamefont {Nam}},
  \bibinfo {author} {\bibfnamefont {Q.}~\bibnamefont {Zhang}}, \bibinfo
  {author} {\bibfnamefont {R.~H.}\ \bibnamefont {Hadfield}}, \bibinfo {author}
  {\bibfnamefont {T.}~\bibnamefont {Honjo}}, \bibinfo {author} {\bibfnamefont
  {K.}~\bibnamefont {Tamaki}}, \ and\ \bibinfo {author} {\bibfnamefont
  {Y.}~\bibnamefont {Yamamoto}},\ }\bibfield  {title} {\enquote {\bibinfo
  {title} {Quantum key distribution over a 40-db channel loss using
  superconducting single-photon detectors},}\ }\href@noop {} {\bibfield
  {journal} {\bibinfo  {journal} {Nat. Photonics}\ }\textbf {\bibinfo {volume}
  {1}},\ \bibinfo {pages} {343} (\bibinfo {year} {2007})}\BibitemShut {NoStop}%
\bibitem [{\citenamefont {van~den Berg}\ \emph {et~al.}(2015)\citenamefont
  {van~den Berg}, \citenamefont {Brange},\ and\ \citenamefont
  {Samuelsson}}]{Ber15}%
  \BibitemOpen
  \bibfield  {author} {\bibinfo {author} {\bibfnamefont {T.~L.}\ \bibnamefont
  {van~den Berg}}, \bibinfo {author} {\bibfnamefont {F.}~\bibnamefont
  {Brange}}, \ and\ \bibinfo {author} {\bibfnamefont {P.}~\bibnamefont
  {Samuelsson}},\ }\bibfield  {title} {\enquote {\bibinfo {title} {Energy and
  temperature fluctuations in the single electron box},}\ }\href
  {http://stacks.iop.org/1367-2630/17/i=7/a=075012} {\bibfield  {journal}
  {\bibinfo  {journal} {New J. Phys.}\ }\textbf {\bibinfo {volume} {17}},\
  \bibinfo {pages} {075012} (\bibinfo {year} {2015})}\BibitemShut {NoStop}%
\bibitem [{\citenamefont {Pekola}(2015)}]{Pek15}%
  \BibitemOpen
  \bibfield  {author} {\bibinfo {author} {\bibfnamefont {J.~P}\ \bibnamefont
  {Pekola}},\ }\bibfield  {title} {\enquote {\bibinfo {title} {Towards quantum
  thermodynamics in electronic circuits},}\ }\href@noop {} {\bibfield
  {journal} {\bibinfo  {journal} {Nat. Phys.}\ }\textbf {\bibinfo {volume}
  {11}},\ \bibinfo {pages} {118} (\bibinfo {year} {2015})}\BibitemShut
  {NoStop}%
\bibitem [{\citenamefont {Marchegiani}\ \emph {et~al.}(2016)\citenamefont
  {Marchegiani}, \citenamefont {Virtanen}, \citenamefont {Giazotto},\ and\
  \citenamefont {Campisi}}]{Mar16}%
  \BibitemOpen
  \bibfield  {author} {\bibinfo {author} {\bibfnamefont {G.}~\bibnamefont
  {Marchegiani}}, \bibinfo {author} {\bibfnamefont {P.}~\bibnamefont
  {Virtanen}}, \bibinfo {author} {\bibfnamefont {F.}~\bibnamefont {Giazotto}},
  \ and\ \bibinfo {author} {\bibfnamefont {M.}~\bibnamefont {Campisi}},\
  }\bibfield  {title} {\enquote {\bibinfo {title} {Self-oscillating Josephson
  quantum heat engine},}\ }\href {\doibase 10.1103/PhysRevApplied.6.054014}
  {\bibfield  {journal} {\bibinfo  {journal} {Phys. Rev. Applied}\ }\textbf
  {\bibinfo {volume} {6}},\ \bibinfo {pages} {054014} (\bibinfo {year}
  {2016})}\BibitemShut {NoStop}%
\bibitem [{\citenamefont {Marchegiani}\ \emph {et~al.}(2018)\citenamefont
  {Marchegiani}, \citenamefont {Virtanen},\ and\ \citenamefont
  {Giazotto}}]{Mar18}%
  \BibitemOpen
  \bibfield  {author} {\bibinfo {author} {\bibfnamefont {G.}~\bibnamefont
  {Marchegiani}}, \bibinfo {author} {\bibfnamefont {P.}~\bibnamefont
  {Virtanen}}, \ and\ \bibinfo {author} {\bibfnamefont {F.}~\bibnamefont
  {Giazotto}},\ }\bibfield  {title} {\enquote {\bibinfo {title} {On-chip
  cooling by heating with superconducting tunnel junctions},}\ }\href
  {http://stacks.iop.org/0295-5075/124/i=4/a=48005} {\bibfield  {journal}
  {\bibinfo  {journal} {EPL (Europhysics Letters)}\ }\textbf {\bibinfo {volume}
  {124}},\ \bibinfo {pages} {48005} (\bibinfo {year} {2018})}\BibitemShut
  {NoStop}%
\bibitem [{\citenamefont {Vischi}\ \emph {et~al.}(2019)\citenamefont {Vischi},
  \citenamefont {Carrega}, \citenamefont {Virtanen}, \citenamefont {Strambini},
  \citenamefont {Braggio},\ and\ \citenamefont {Giazotto}}]{Vis18}%
  \BibitemOpen
  \bibfield  {author} {\bibinfo {author} {\bibfnamefont {F.}~\bibnamefont
  {Vischi}}, \bibinfo {author} {\bibfnamefont {M.}~\bibnamefont {Carrega}},
  \bibinfo {author} {\bibfnamefont {P.}~\bibnamefont {Virtanen}}, \bibinfo
  {author} {\bibfnamefont {E.}~\bibnamefont {Strambini}}, \bibinfo {author}
  {\bibfnamefont {A.}~\bibnamefont {Braggio}}, \ and\ \bibinfo {author}
  {\bibfnamefont {F.}~\bibnamefont {Giazotto}},\ }\bibfield  {title} {\enquote
  {\bibinfo {title} {Thermodynamic cycles in Josephson junctions},}\ }\href
  {\doibase 10.1038/s41598-019-40202-8} {\bibfield  {journal} {\bibinfo
  {journal} {Sci. Rep.}\ }\textbf {\bibinfo {volume} {9}},\ \bibinfo {pages}
  {3238} (\bibinfo {year} {2019})}\BibitemShut {NoStop}%
\bibitem [{\citenamefont {Guarcello}\ \emph {et~al.}(2019)\citenamefont
  {Guarcello}, \citenamefont {Braggio}, \citenamefont {Solinas},\ and\
  \citenamefont {Giazotto}}]{GuaBra18}%
  \BibitemOpen
  \bibfield  {author} {\bibinfo {author} {\bibfnamefont {C}~\bibnamefont
  {Guarcello}}, \bibinfo {author} {\bibfnamefont {A}~\bibnamefont {Braggio}},
  \bibinfo {author} {\bibfnamefont {P}~\bibnamefont {Solinas}}, \ and\ \bibinfo
  {author} {\bibfnamefont {F}~\bibnamefont {Giazotto}},\ }\bibfield  {title}
  {\enquote {\bibinfo {title} {Nonlinear critical-current thermal response of
  an asymmetric Josephson tunnel junction},}\ }\href {\doibase
  10.1103/PhysRevApplied.11.024002} {\bibfield  {journal} {\bibinfo  {journal}
  {Phys. Rev. Applied}\ }\textbf {\bibinfo {volume} {11}},\ \bibinfo {pages}
  {024002} (\bibinfo {year} {2019})}\BibitemShut {NoStop}%
\bibitem [{\citenamefont {Mart{\'i}nez-P{\'e}rez}\ \emph
  {et~al.}(2014)\citenamefont {Mart{\'i}nez-P{\'e}rez}, \citenamefont
  {Solinas},\ and\ \citenamefont {Giazotto}}]{MarSol14}%
  \BibitemOpen
  \bibfield  {author} {\bibinfo {author} {\bibfnamefont {M.~J.}\ \bibnamefont
  {Mart{\'i}nez-P{\'e}rez}}, \bibinfo {author} {\bibfnamefont {P.}~\bibnamefont
  {Solinas}}, \ and\ \bibinfo {author} {\bibfnamefont {F.}~\bibnamefont
  {Giazotto}},\ }\bibfield  {title} {\enquote {\bibinfo {title} {Coherent
  caloritronics in Josephson-based nanocircuits},}\ }\href {\doibase
  10.1007/s10909-014-1132-6} {\bibfield  {journal} {\bibinfo  {journal} {J. Low
  Temp. Phys.}\ }\textbf {\bibinfo {volume} {175}},\ \bibinfo {pages}
  {813--837} (\bibinfo {year} {2014})}\BibitemShut {NoStop}%
\bibitem [{\citenamefont {Fornieri}\ and\ \citenamefont
  {Giazotto}(2017)}]{ForGia17}%
  \BibitemOpen
  \bibfield  {author} {\bibinfo {author} {\bibfnamefont {A.}~\bibnamefont
  {Fornieri}}\ and\ \bibinfo {author} {\bibfnamefont {F.}~\bibnamefont
  {Giazotto}},\ }\bibfield  {title} {\enquote {\bibinfo {title} {Towards
  phase-coherent caloritronics in superconducting circuits},}\ }\href@noop {}
  {\bibfield  {journal} {\bibinfo  {journal} {Nat. Nanotechnology}\ }\textbf
  {\bibinfo {volume} {12}},\ \bibinfo {pages} {944--952} (\bibinfo {year}
  {2017})}\BibitemShut {NoStop}%
\bibitem [{\citenamefont {Giazotto}\ and\ \citenamefont
  {Mart{\'\i}nez-P{\'e}rez}(2012)}]{Gia12}%
  \BibitemOpen
  \bibfield  {author} {\bibinfo {author} {\bibfnamefont {F.}~\bibnamefont
  {Giazotto}}\ and\ \bibinfo {author} {\bibfnamefont {M.~J.}\ \bibnamefont
  {Mart{\'\i}nez-P{\'e}rez}},\ }\bibfield  {title} {\enquote {\bibinfo {title}
  {The Josephson heat interferometer},}\ }\href@noop {} {\bibfield  {journal}
  {\bibinfo  {journal} {Nature}\ }\textbf {\bibinfo {volume} {492}},\ \bibinfo
  {pages} {401--405} (\bibinfo {year} {2012})}\BibitemShut {NoStop}%
\bibitem [{\citenamefont {Mart{\'\i}nez-P{\'e}rez}\ and\ \citenamefont
  {Giazotto}(2014)}]{Mar14}%
  \BibitemOpen
  \bibfield  {author} {\bibinfo {author} {\bibfnamefont {M.~J.}\
  \bibnamefont {Mart{\'\i}nez-P{\'e}rez}}\ and\ \bibinfo {author}
  {\bibfnamefont {F.}\ \bibnamefont {Giazotto}},\ }\bibfield  {title}
  {\enquote {\bibinfo {title} {A quantum diffractor for thermal flux},}\
  }\href@noop {} {\bibfield  {journal} {\bibinfo  {journal} {Nat. Commun.}\
  }\textbf {\bibinfo {volume} {5}},\ \bibinfo {pages} {3579} (\bibinfo {year}
  {2014})}\BibitemShut {NoStop}%
\bibitem [{\citenamefont {Guarcello}\ \emph {et~al.}(2016)\citenamefont
  {Guarcello}, \citenamefont {Giazotto},\ and\ \citenamefont
  {Solinas}}]{Gua16}%
  \BibitemOpen
  \bibfield  {author} {\bibinfo {author} {\bibfnamefont {C.}~\bibnamefont
  {Guarcello}}, \bibinfo {author} {\bibfnamefont {F.}~\bibnamefont {Giazotto}},
  \ and\ \bibinfo {author} {\bibfnamefont {P.}~\bibnamefont {Solinas}},\
  }\bibfield  {title} {\enquote {\bibinfo {title} {Coherent diffraction of
  thermal currents in long Josephson tunnel junctions},}\ }\href {\doibase
  10.1103/PhysRevB.94.054522} {\bibfield  {journal} {\bibinfo  {journal} {Phys.
  Rev. B}\ }\textbf {\bibinfo {volume} {94}},\ \bibinfo {pages} {054522}
  (\bibinfo {year} {2016})}\BibitemShut {NoStop}%
\bibitem [{\citenamefont {Mart{\'\i}nez-P{\'e}rez}\ \emph
  {et~al.}(2015)\citenamefont {Mart{\'\i}nez-P{\'e}rez}, \citenamefont
  {Fornieri},\ and\ \citenamefont {Giazotto}}]{Mar15}%
  \BibitemOpen
  \bibfield  {author} {\bibinfo {author} {\bibfnamefont {M.~J.}\ \bibnamefont
  {Mart{\'\i}nez-P{\'e}rez}}, \bibinfo {author} {\bibfnamefont
  {A.}~\bibnamefont {Fornieri}}, \ and\ \bibinfo {author} {\bibfnamefont
  {F.}~\bibnamefont {Giazotto}},\ }\bibfield  {title} {\enquote {\bibinfo
  {title} {Rectification of electronic heat current by a hybrid thermal
  diode},}\ }\href@noop {} {\bibfield  {journal} {\bibinfo  {journal} {Nature
  Nanotechnology}\ }\textbf {\bibinfo {volume} {10}},\ \bibinfo {pages}
  {303--307} (\bibinfo {year} {2015})}\BibitemShut {NoStop}%
\bibitem [{\citenamefont {Fornieri}\ \emph {et~al.}(2016)\citenamefont
  {Fornieri}, \citenamefont {Timossi}, \citenamefont {Bosisio}, \citenamefont
  {Solinas},\ and\ \citenamefont {Giazotto}}]{For16}%
  \BibitemOpen
  \bibfield  {author} {\bibinfo {author} {\bibfnamefont {A.}~\bibnamefont
  {Fornieri}}, \bibinfo {author} {\bibfnamefont {G.}~\bibnamefont {Timossi}},
  \bibinfo {author} {\bibfnamefont {R.}~\bibnamefont {Bosisio}}, \bibinfo
  {author} {\bibfnamefont {P.}~\bibnamefont {Solinas}}, \ and\ \bibinfo
  {author} {\bibfnamefont {F.}~\bibnamefont {Giazotto}},\ }\bibfield  {title}
  {\enquote {\bibinfo {title} {Negative differential thermal conductance and
  heat amplification in superconducting hybrid devices},}\ }\href {\doibase
  10.1103/PhysRevB.93.134508} {\bibfield  {journal} {\bibinfo  {journal} {Phys.
  Rev. B}\ }\textbf {\bibinfo {volume} {93}},\ \bibinfo {pages} {134508}
  (\bibinfo {year} {2016})}\BibitemShut {NoStop}%
\bibitem [{\citenamefont {Guarcello}\ \emph
  {et~al.}(2017{\natexlab{a}})\citenamefont {Guarcello}, \citenamefont
  {Solinas}, \citenamefont {Di~Ventra},\ and\ \citenamefont
  {Giazotto}}]{Gua17}%
  \BibitemOpen
  \bibfield  {author} {\bibinfo {author} {\bibfnamefont {C.}~\bibnamefont
  {Guarcello}}, \bibinfo {author} {\bibfnamefont {P.}~\bibnamefont {Solinas}},
  \bibinfo {author} {\bibfnamefont {M.}~\bibnamefont {Di~Ventra}}, \ and\
  \bibinfo {author} {\bibfnamefont {F.}~\bibnamefont {Giazotto}},\ }\bibfield
  {title} {\enquote {\bibinfo {title} {Hysteretic superconducting heat-flux
  quantum modulator},}\ }\href {\doibase 10.1103/PhysRevApplied.7.044021}
  {\bibfield  {journal} {\bibinfo  {journal} {Phys. Rev. Applied}\ }\textbf
  {\bibinfo {volume} {7}},\ \bibinfo {pages} {044021} (\bibinfo {year}
  {2017}{\natexlab{a}})}\BibitemShut {NoStop}%
\bibitem [{\citenamefont {Guarcello}\ \emph
  {et~al.}(2017{\natexlab{b}})\citenamefont {Guarcello}, \citenamefont
  {Solinas}, \citenamefont {Di~Ventra},\ and\ \citenamefont
  {Giazotto}}]{GuaSol17}%
  \BibitemOpen
  \bibfield  {author} {\bibinfo {author} {\bibfnamefont {C.}~\bibnamefont
  {Guarcello}}, \bibinfo {author} {\bibfnamefont {P.}~\bibnamefont {Solinas}},
  \bibinfo {author} {\bibfnamefont {M.}~\bibnamefont {Di~Ventra}}, \ and\
  \bibinfo {author} {\bibfnamefont {F.}~\bibnamefont {Giazotto}},\ }\bibfield
  {title} {\enquote {\bibinfo {title} {Solitonic Josephson-based meminductive
  systems},}\ }\href@noop {} {\bibfield  {journal} {\bibinfo  {journal} {Sci.
  Rep.}\ }\textbf {\bibinfo {volume} {7}},\ \bibinfo {pages} {46736} (\bibinfo
  {year} {2017}{\natexlab{b}})}\BibitemShut {NoStop}%
\bibitem [{\citenamefont {Guarcello}\ \emph
  {et~al.}(2018{\natexlab{a}})\citenamefont {Guarcello}, \citenamefont
  {Solinas}, \citenamefont {Braggio}, \citenamefont {Di~Ventra},\ and\
  \citenamefont {Giazotto}}]{GuaSol18}%
  \BibitemOpen
  \bibfield  {author} {\bibinfo {author} {\bibfnamefont {C.}~\bibnamefont
  {Guarcello}}, \bibinfo {author} {\bibfnamefont {P.}~\bibnamefont {Solinas}},
  \bibinfo {author} {\bibfnamefont {A.}~\bibnamefont {Braggio}}, \bibinfo
  {author} {\bibfnamefont {M.}~\bibnamefont {Di~Ventra}}, \ and\ \bibinfo
  {author} {\bibfnamefont {F.}~\bibnamefont {Giazotto}},\ }\bibfield  {title}
  {\enquote {\bibinfo {title} {Josephson thermal memory},}\ }\href {\doibase
  10.1103/PhysRevApplied.9.014021} {\bibfield  {journal} {\bibinfo  {journal}
  {Phys. Rev. Applied}\ }\textbf {\bibinfo {volume} {9}},\ \bibinfo {pages}
  {014021} (\bibinfo {year} {2018}{\natexlab{a}})}\BibitemShut {NoStop}%
\bibitem [{\citenamefont {Paolucci}\ \emph {et~al.}(2018)\citenamefont
  {Paolucci}, \citenamefont {Marchegiani}, \citenamefont {Strambini},\ and\
  \citenamefont {Giazotto}}]{Pao18}%
  \BibitemOpen
  \bibfield  {author} {\bibinfo {author} {\bibfnamefont {F.}~\bibnamefont
  {Paolucci}}, \bibinfo {author} {\bibfnamefont {G.}~\bibnamefont
  {Marchegiani}}, \bibinfo {author} {\bibfnamefont {E.}~\bibnamefont
  {Strambini}}, \ and\ \bibinfo {author} {\bibfnamefont {F.}~\bibnamefont
  {Giazotto}},\ }\bibfield  {title} {\enquote {\bibinfo {title} {Phase-tunable
  thermal logic: Computation with heat},}\ }\href {\doibase
  10.1103/PhysRevApplied.10.024003} {\bibfield  {journal} {\bibinfo  {journal}
  {Phys. Rev. Applied}\ }\textbf {\bibinfo {volume} {10}},\ \bibinfo {pages}
  {024003} (\bibinfo {year} {2018})}\BibitemShut {NoStop}%
\bibitem [{\citenamefont {Sothmann}\ \emph {et~al.}(2017)\citenamefont
  {Sothmann}, \citenamefont {Giazotto},\ and\ \citenamefont
  {Hankiewicz}}]{Sot17}%
  \BibitemOpen
  \bibfield  {author} {\bibinfo {author} {\bibfnamefont {B.}~\bibnamefont
  {Sothmann}}, \bibinfo {author} {\bibfnamefont {F.}~\bibnamefont {Giazotto}},
  \ and\ \bibinfo {author} {\bibfnamefont {E.~M}\ \bibnamefont {Hankiewicz}},\
  }\bibfield  {title} {\enquote {\bibinfo {title} {High-efficiency thermal
  switch based on topological Josephson junctions},}\ }\href
  {http://stacks.iop.org/1367-2630/19/i=2/a=023056} {\bibfield  {journal}
  {\bibinfo  {journal} {New Journal of Physics}\ }\textbf {\bibinfo {volume}
  {19}},\ \bibinfo {pages} {023056} (\bibinfo {year} {2017})}\BibitemShut
  {NoStop}%
\bibitem [{\citenamefont {Timossi}\ \emph {et~al.}(2018)\citenamefont
  {Timossi}, \citenamefont {Fornieri}, \citenamefont {Paolucci}, \citenamefont
  {Puglia},\ and\ \citenamefont {Giazotto}}]{Tim18}%
  \BibitemOpen
  \bibfield  {author} {\bibinfo {author} {\bibfnamefont {G.~F.}\ \bibnamefont
  {Timossi}}, \bibinfo {author} {\bibfnamefont {A.}~\bibnamefont {Fornieri}},
  \bibinfo {author} {\bibfnamefont {F.}~\bibnamefont {Paolucci}}, \bibinfo
  {author} {\bibfnamefont {C.}~\bibnamefont {Puglia}}, \ and\ \bibinfo {author}
  {\bibfnamefont {F.}~\bibnamefont {Giazotto}},\ }\bibfield  {title} {\enquote
  {\bibinfo {title} {Phase-tunable Josephson thermal router},}\ }\href
  {\doibase 10.1021/acs.nanolett.7b04906} {\bibfield  {journal} {\bibinfo
  {journal} {Nano Letters}\ }\textbf {\bibinfo {volume} {18}},\ \bibinfo
  {pages} {1764--1769} (\bibinfo {year} {2018})}\BibitemShut {NoStop}%
\bibitem [{\citenamefont {Guarcello}\ \emph
  {et~al.}(2018{\natexlab{b}})\citenamefont {Guarcello}, \citenamefont
  {Solinas}, \citenamefont {Braggio},\ and\ \citenamefont {Giazotto}}]{Gua18}%
  \BibitemOpen
  \bibfield  {author} {\bibinfo {author} {\bibfnamefont {C.}~\bibnamefont
  {Guarcello}}, \bibinfo {author} {\bibfnamefont {P.}~\bibnamefont {Solinas}},
  \bibinfo {author} {\bibfnamefont {A.}~\bibnamefont {Braggio}}, \ and\
  \bibinfo {author} {\bibfnamefont {F.}~\bibnamefont {Giazotto}},\ }\bibfield
  {title} {\enquote {\bibinfo {title} {Solitonic Josephson thermal
  transport},}\ }\href {\doibase 10.1103/PhysRevApplied.9.034014} {\bibfield
  {journal} {\bibinfo  {journal} {Phys. Rev. Applied}\ }\textbf {\bibinfo
  {volume} {9}},\ \bibinfo {pages} {034014} (\bibinfo {year}
  {2018}{\natexlab{b}})}\BibitemShut {NoStop}%
\bibitem [{\citenamefont {Hwang}\ \emph {et~al.}(2018)\citenamefont {Hwang},
  \citenamefont {Giazotto},\ and\ \citenamefont {Sothmann}}]{Hwa18}%
  \BibitemOpen
  \bibfield  {author} {\bibinfo {author} {\bibfnamefont {S.-Y.}\ \bibnamefont
  {Hwang}}, \bibinfo {author} {\bibfnamefont {F.}~\bibnamefont {Giazotto}}, \
  and\ \bibinfo {author} {\bibfnamefont {B.}~\bibnamefont {Sothmann}},\
  }\bibfield  {title} {\enquote {\bibinfo {title} {Phase-coherent heat
  circulator based on multiterminal Josephson junctions},}\ }\href {\doibase
  10.1103/PhysRevApplied.10.044062} {\bibfield  {journal} {\bibinfo  {journal}
  {Phys. Rev. Applied}\ }\textbf {\bibinfo {volume} {10}},\ \bibinfo {pages}
  {044062} (\bibinfo {year} {2018})}\BibitemShut {NoStop}%
\bibitem [{\citenamefont {Wellstood}\ \emph {et~al.}(1994)\citenamefont
  {Wellstood}, \citenamefont {Urbina},\ and\ \citenamefont {Clarke}}]{Wel94}%
  \BibitemOpen
  \bibfield  {author} {\bibinfo {author} {\bibfnamefont {F.~C.}\ \bibnamefont
  {Wellstood}}, \bibinfo {author} {\bibfnamefont {C.}~\bibnamefont {Urbina}}, \
  and\ \bibinfo {author} {\bibfnamefont {John}\ \bibnamefont {Clarke}},\
  }\bibfield  {title} {\enquote {\bibinfo {title} {Hot-electron effects in
  metals},}\ }\href {\doibase 10.1103/PhysRevB.49.5942} {\bibfield  {journal}
  {\bibinfo  {journal} {Phys. Rev. B}\ }\textbf {\bibinfo {volume} {49}},\
  \bibinfo {pages} {5942--5955} (\bibinfo {year} {1994})}\BibitemShut {NoStop}%
\bibitem [{\citenamefont {Tinkham}(2004)}]{Tin04}%
  \BibitemOpen
  \bibfield  {author} {\bibinfo {author} {\bibfnamefont {M.}~\bibnamefont
  {Tinkham}},\ }\href@noop {} {\emph {\bibinfo {title} {Introduction to
  Superconductivity}}},\ Dover Books on Physics Series\ (\bibinfo  {publisher}
  {Dover Publications},\ \bibinfo {year} {2004})\BibitemShut {NoStop}%
\bibitem [{\citenamefont {Golubov}\ \emph {et~al.}(2004)\citenamefont
  {Golubov}, \citenamefont {Kupriyanov},\ and\ \citenamefont
  {Il'ichev}}]{Gol04}%
  \BibitemOpen
  \bibfield  {author} {\bibinfo {author} {\bibfnamefont {A.~A.}\ \bibnamefont
  {Golubov}}, \bibinfo {author} {\bibfnamefont {M.~Yu.}\ \bibnamefont
  {Kupriyanov}}, \ and\ \bibinfo {author} {\bibfnamefont {E.}~\bibnamefont
  {Il'ichev}},\ }\bibfield  {title} {\enquote {\bibinfo {title} {The
  current-phase relation in Josephson junctions},}\ }\href {\doibase
  10.1103/RevModPhys.76.411} {\bibfield  {journal} {\bibinfo  {journal} {Rev.
  Mod. Phys.}\ }\textbf {\bibinfo {volume} {76}},\ \bibinfo {pages} {411--469}
  (\bibinfo {year} {2004})}\BibitemShut {NoStop}%
\bibitem [{\citenamefont {Giazotto}\ and\ \citenamefont
  {Pekola}(2005)}]{Gia05}%
  \BibitemOpen
  \bibfield  {author} {\bibinfo {author} {\bibfnamefont {F.}~\bibnamefont
  {Giazotto}}\ and\ \bibinfo {author} {\bibfnamefont {J.~P.}\ \bibnamefont
  {Pekola}},\ }\bibfield  {title} {\enquote {\bibinfo {title} {Josephson tunnel
  junction controlled by quasiparticle injection},}\ }\href@noop {} {\bibfield
  {journal} {\bibinfo  {journal} {J. Appl. Phys.}\ }\textbf {\bibinfo {volume}
  {97}},\ \bibinfo {eid} {023908} (\bibinfo {year} {2005})}\BibitemShut
  {NoStop}%
\bibitem [{\citenamefont {Tirelli}\ \emph {et~al.}(2008)\citenamefont
  {Tirelli}, \citenamefont {Savin}, \citenamefont {Garcia}, \citenamefont
  {Pekola}, \citenamefont {Beltram},\ and\ \citenamefont {Giazotto}}]{Tir08}%
  \BibitemOpen
  \bibfield  {author} {\bibinfo {author} {\bibfnamefont {S.}~\bibnamefont
  {Tirelli}}, \bibinfo {author} {\bibfnamefont {A.~M.}\ \bibnamefont {Savin}},
  \bibinfo {author} {\bibfnamefont {C.~Pascual}\ \bibnamefont {Garcia}},
  \bibinfo {author} {\bibfnamefont {J.~P.}\ \bibnamefont {Pekola}}, \bibinfo
  {author} {\bibfnamefont {F.}~\bibnamefont {Beltram}}, \ and\ \bibinfo
  {author} {\bibfnamefont {F.}~\bibnamefont {Giazotto}},\ }\bibfield  {title}
  {\enquote {\bibinfo {title} {Manipulation and generation of supercurrent in
  out-of-equilibrium Josephson tunnel nanojunctions},}\ }\href {\doibase
  10.1103/PhysRevLett.101.077004} {\bibfield  {journal} {\bibinfo  {journal}
  {Phys. Rev. Lett.}\ }\textbf {\bibinfo {volume} {101}},\ \bibinfo {pages}
  {077004} (\bibinfo {year} {2008})}\BibitemShut {NoStop}%
\bibitem [{\citenamefont {Barone}\ and\ \citenamefont
  {Patern\`{o}}(1982)}]{Bar82}%
  \BibitemOpen
  \bibfield  {author} {\bibinfo {author} {\bibfnamefont {A.}~\bibnamefont
  {Barone}}\ and\ \bibinfo {author} {\bibfnamefont {G.}~\bibnamefont
  {Patern\`{o}}},\ }\href@noop {} {\emph {\bibinfo {title} {Physics and
  Applications of the Josephson Effect}}}\ (\bibinfo  {publisher} {Wiley, New
  York},\ \bibinfo {year} {1982})\BibitemShut {NoStop}%
\bibitem [{\citenamefont {Dynes}\ \emph {et~al.}(1978)\citenamefont {Dynes},
  \citenamefont {Narayanamurti},\ and\ \citenamefont {Garno}}]{Dyn78}%
  \BibitemOpen
  \bibfield  {author} {\bibinfo {author} {\bibfnamefont {R.~C.}\ \bibnamefont
  {Dynes}}, \bibinfo {author} {\bibfnamefont {V.}~\bibnamefont
  {Narayanamurti}}, \ and\ \bibinfo {author} {\bibfnamefont {J.~P.}\
  \bibnamefont {Garno}},\ }\bibfield  {title} {\enquote {\bibinfo {title}
  {Direct measurement of quasiparticle-lifetime broadening in a strong-coupled
  superconductor},}\ }\href {\doibase 10.1103/PhysRevLett.41.1509} {\bibfield
  {journal} {\bibinfo  {journal} {Phys. Rev. Lett.}\ }\textbf {\bibinfo
  {volume} {41}},\ \bibinfo {pages} {1509--1512} (\bibinfo {year}
  {1978})}\BibitemShut {NoStop}%
\bibitem [{\citenamefont {Pekola}\ \emph {et~al.}(2010)\citenamefont {Pekola},
  \citenamefont {Maisi}, \citenamefont {Kafanov}, \citenamefont {Chekurov},
  \citenamefont {Kemppinen}, \citenamefont {Pashkin}, \citenamefont {Saira},
  \citenamefont {M\"ott\"onen},\ and\ \citenamefont {Tsai}}]{Pek10}%
  \BibitemOpen
  \bibfield  {author} {\bibinfo {author} {\bibfnamefont {J.~P.}\ \bibnamefont
  {Pekola}}, \bibinfo {author} {\bibfnamefont {V.~F.}\ \bibnamefont {Maisi}},
  \bibinfo {author} {\bibfnamefont {S.}~\bibnamefont {Kafanov}}, \bibinfo
  {author} {\bibfnamefont {N.}~\bibnamefont {Chekurov}}, \bibinfo {author}
  {\bibfnamefont {A.}~\bibnamefont {Kemppinen}}, \bibinfo {author}
  {\bibfnamefont {Yu.~A.}\ \bibnamefont {Pashkin}}, \bibinfo {author}
  {\bibfnamefont {O.-P.}\ \bibnamefont {Saira}}, \bibinfo {author}
  {\bibfnamefont {M.}~\bibnamefont {M\"ott\"onen}}, \ and\ \bibinfo {author}
  {\bibfnamefont {J.~S.}\ \bibnamefont {Tsai}},\ }\bibfield  {title} {\enquote
  {\bibinfo {title} {Environment-assisted tunneling as an origin of the Dynes
  density of states},}\ }\href {\doibase 10.1103/PhysRevLett.105.026803}
  {\bibfield  {journal} {\bibinfo  {journal} {Phys. Rev. Lett.}\ }\textbf
  {\bibinfo {volume} {105}},\ \bibinfo {pages} {026803} (\bibinfo {year}
  {2010})}\BibitemShut {NoStop}%
\bibitem [{\citenamefont {Clarke}\ and\ \citenamefont
  {Braginski}(2004)}]{Cla04}%
  \BibitemOpen
  \bibfield  {author} {\bibinfo {author} {\bibfnamefont {J.}~\bibnamefont
  {Clarke}}\ and\ \bibinfo {author} {\bibfnamefont {A.I.}\ \bibnamefont
  {Braginski}},\ }\href@noop {} {\emph {\bibinfo {title} {The SQUID Handbook:
  Fundamentals and Technology of SQUIDs and SQUID Systems}}},\ \bibinfo
  {series} {The SQUID Handbook}\ No.\ \bibinfo {number} {v. 1}\ (\bibinfo
  {publisher} {Wiley},\ \bibinfo {year} {2004})\BibitemShut {NoStop}%
\bibitem [{\citenamefont {Andreev}(1964)}]{And64}%
  \BibitemOpen
  \bibfield  {author} {\bibinfo {author} {\bibfnamefont {A.~F.}\ \bibnamefont
  {Andreev}},\ }\bibfield  {title} {\enquote {\bibinfo {title} {The thermal
  conductivity of the intermediate state in superconductors},}\ }\href@noop {}
  {\bibfield  {journal} {\bibinfo  {journal} {J. Exp. Theor. Phys.}\ }\textbf
  {\bibinfo {volume} {19}},\ \bibinfo {pages} {1228} (\bibinfo {year}
  {1964})}\BibitemShut {NoStop}%
\bibitem [{\citenamefont {Moseley}\ \emph {et~al.}(1984)\citenamefont
  {Moseley}, \citenamefont {Mather},\ and\ \citenamefont {McCammon}}]{Mos84}%
  \BibitemOpen
  \bibfield  {author} {\bibinfo {author} {\bibfnamefont {S.~H.}\ \bibnamefont
  {Moseley}}, \bibinfo {author} {\bibfnamefont {J.~C.}\ \bibnamefont {Mather}},
  \ and\ \bibinfo {author} {\bibfnamefont {D.}~\bibnamefont {McCammon}},\
  }\bibfield  {title} {\enquote {\bibinfo {title} {Thermal detectors as x--ray
  spectrometers},}\ }\href {\doibase 10.1063/1.334129} {\bibfield  {journal}
  {\bibinfo  {journal} {J. Appl. Phys.}\ }\textbf {\bibinfo {volume} {56}},\
  \bibinfo {pages} {1257--1262} (\bibinfo {year} {1984})}\BibitemShut {NoStop}%
\bibitem [{\citenamefont {Chui}\ \emph {et~al.}(1992)\citenamefont {Chui},
  \citenamefont {Swanson}, \citenamefont {Adriaans}, \citenamefont {Nissen},\
  and\ \citenamefont {Lipa}}]{Chu92}%
  \BibitemOpen
  \bibfield  {author} {\bibinfo {author} {\bibfnamefont {T.~C.~P.}\
  \bibnamefont {Chui}}, \bibinfo {author} {\bibfnamefont {D.~R.}\ \bibnamefont
  {Swanson}}, \bibinfo {author} {\bibfnamefont {M.~J.}\ \bibnamefont
  {Adriaans}}, \bibinfo {author} {\bibfnamefont {J.~A.}\ \bibnamefont
  {Nissen}}, \ and\ \bibinfo {author} {\bibfnamefont {J.~A.}\ \bibnamefont
  {Lipa}},\ }\bibfield  {title} {\enquote {\bibinfo {title} {Temperature
  fluctuations in the canonical ensemble},}\ }\href {\doibase
  10.1103/PhysRevLett.69.3005} {\bibfield  {journal} {\bibinfo  {journal}
  {Phys. Rev. Lett.}\ }\textbf {\bibinfo {volume} {69}},\ \bibinfo {pages}
  {3005--3008} (\bibinfo {year} {1992})}\BibitemShut {NoStop}%
\bibitem [{\citenamefont {Rabani}\ \emph {et~al.}(2008)\citenamefont {Rabani},
  \citenamefont {Taddei}, \citenamefont {Bourgeois}, \citenamefont {Fazio},\
  and\ \citenamefont {Giazotto}}]{Rab08}%
  \BibitemOpen
  \bibfield  {author} {\bibinfo {author} {\bibfnamefont {H.}~\bibnamefont
  {Rabani}}, \bibinfo {author} {\bibfnamefont {F.}~\bibnamefont {Taddei}},
  \bibinfo {author} {\bibfnamefont {O.}~\bibnamefont {Bourgeois}}, \bibinfo
  {author} {\bibfnamefont {R.}~\bibnamefont {Fazio}}, \ and\ \bibinfo {author}
  {\bibfnamefont {F.}~\bibnamefont {Giazotto}},\ }\bibfield  {title} {\enquote
  {\bibinfo {title} {Phase-dependent electronic specific heat of mesoscopic
  Josephson junctions},}\ }\href {\doibase 10.1103/PhysRevB.78.012503}
  {\bibfield  {journal} {\bibinfo  {journal} {Phys. Rev. B}\ }\textbf {\bibinfo
  {volume} {78}},\ \bibinfo {pages} {012503} (\bibinfo {year}
  {2008})}\BibitemShut {NoStop}%
\bibitem [{\citenamefont {Solinas}\ \emph {et~al.}(2016)\citenamefont
  {Solinas}, \citenamefont {Bosisio},\ and\ \citenamefont {Giazotto}}]{Sol16}%
  \BibitemOpen
  \bibfield  {author} {\bibinfo {author} {\bibfnamefont {P.}~\bibnamefont
  {Solinas}}, \bibinfo {author} {\bibfnamefont {R.}~\bibnamefont {Bosisio}}, \
  and\ \bibinfo {author} {\bibfnamefont {F.}~\bibnamefont {Giazotto}},\
  }\bibfield  {title} {\enquote {\bibinfo {title} {Microwave quantum
  refrigeration based on the Josephson effect},}\ }\href {\doibase
  10.1103/PhysRevB.93.224521} {\bibfield  {journal} {\bibinfo  {journal} {Phys.
  Rev. B}\ }\textbf {\bibinfo {volume} {93}},\ \bibinfo {pages} {224521}
  (\bibinfo {year} {2016})}\BibitemShut {NoStop}%
\bibitem [{\citenamefont {Walsh}\ \emph {et~al.}(2017)\citenamefont {Walsh},
  \citenamefont {Efetov}, \citenamefont {Lee}, \citenamefont {Heuck},
  \citenamefont {Crossno}, \citenamefont {Ohki}, \citenamefont {Kim},
  \citenamefont {Englund},\ and\ \citenamefont {Fong}}]{Wal17}%
  \BibitemOpen
  \bibfield  {author} {\bibinfo {author} {\bibfnamefont {E.~D.}\ \bibnamefont
  {Walsh}}, \bibinfo {author} {\bibfnamefont {D.~K.}\ \bibnamefont {Efetov}},
  \bibinfo {author} {\bibfnamefont {G.-H.}\ \bibnamefont {Lee}}, \bibinfo
  {author} {\bibfnamefont {M.}~\bibnamefont {Heuck}}, \bibinfo {author}
  {\bibfnamefont {J.}~\bibnamefont {Crossno}}, \bibinfo {author} {\bibfnamefont
  {T.~A.}\ \bibnamefont {Ohki}}, \bibinfo {author} {\bibfnamefont
  {P.}~\bibnamefont {Kim}}, \bibinfo {author} {\bibfnamefont {D.}~\bibnamefont
  {Englund}}, \ and\ \bibinfo {author} {\bibfnamefont {K.~C.}\ \bibnamefont
  {Fong}},\ }\bibfield  {title} {\enquote {\bibinfo {title} {Graphene-based
  Josephson-junction single-photon detector},}\ }\href {\doibase
  10.1103/PhysRevApplied.8.024022} {\bibfield  {journal} {\bibinfo  {journal}
  {Phys. Rev. Applied}\ }\textbf {\bibinfo {volume} {8}},\ \bibinfo {pages}
  {024022} (\bibinfo {year} {2017})}\BibitemShut {NoStop}%
\bibitem [{\citenamefont {Abrikosov}\ \emph {et~al.}(1975)\citenamefont
  {Abrikosov}, \citenamefont {Gorkov},\ and\ \citenamefont
  {Dzyaloshinski}}]{Abr75}%
  \BibitemOpen
  \bibfield  {author} {\bibinfo {author} {\bibfnamefont {A.A.}\ \bibnamefont
  {Abrikosov}}, \bibinfo {author} {\bibfnamefont {L.P.}\ \bibnamefont
  {Gorkov}}, \ and\ \bibinfo {author} {\bibfnamefont {I.E.}\ \bibnamefont
  {Dzyaloshinski}},\ }\href@noop {} {\emph {\bibinfo {title} {Methods of
  Quantum Field Theory in Statistical Physics}}},\ Dover Books on Physics
  Series\ (\bibinfo  {publisher} {Dover Publications},\ \bibinfo {year}
  {1975})\BibitemShut {NoStop}%
\bibitem [{\citenamefont {de~Gennes}(1999)}]{DeG99}%
  \BibitemOpen
  \bibfield  {author} {\bibinfo {author} {\bibfnamefont {P.}~\bibnamefont
  {de~Gennes}},\ }\href@noop {} {\emph {\bibinfo {title} {Superconductivity of
  Metals and Alloys}}},\ Advanced book classics\ (\bibinfo  {publisher}
  {Advanced Book Program, Perseus Books},\ \bibinfo {year} {1999})\BibitemShut
  {NoStop}%
\bibitem [{\citenamefont {Maki}\ and\ \citenamefont {Griffin}(1965)}]{Mak65}%
  \BibitemOpen
  \bibfield  {author} {\bibinfo {author} {\bibfnamefont {K.}~\bibnamefont
  {Maki}}\ and\ \bibinfo {author} {\bibfnamefont {A.}~\bibnamefont {Griffin}},\
  }\bibfield  {title} {\enquote {\bibinfo {title} {Entropy transport between
  two superconductors by electron tunneling},}\ }\href {\doibase
  10.1103/PhysRevLett.15.921} {\bibfield  {journal} {\bibinfo  {journal} {Phys.
  Rev. Lett.}\ }\textbf {\bibinfo {volume} {15}},\ \bibinfo {pages} {921--923}
  (\bibinfo {year} {1965})}\BibitemShut {NoStop}%
\bibitem [{\citenamefont {Golubev}\ \emph {et~al.}(2013)\citenamefont
  {Golubev}, \citenamefont {Faivre},\ and\ \citenamefont {Pekola}}]{Gol13}%
  \BibitemOpen
  \bibfield  {author} {\bibinfo {author} {\bibfnamefont {D.}~\bibnamefont
  {Golubev}}, \bibinfo {author} {\bibfnamefont {T.}~\bibnamefont {Faivre}}, \
  and\ \bibinfo {author} {\bibfnamefont {J.~P.}\ \bibnamefont {Pekola}},\
  }\bibfield  {title} {\enquote {\bibinfo {title} {Heat transport through a
  Josephson junction},}\ }\href {\doibase 10.1103/PhysRevB.87.094522}
  {\bibfield  {journal} {\bibinfo  {journal} {Phys. Rev. B}\ }\textbf {\bibinfo
  {volume} {87}},\ \bibinfo {pages} {094522} (\bibinfo {year}
  {2013})}\BibitemShut {NoStop}%
  \bibitem [{Note2()}]{Note2}%
  \BibitemOpen
  \bibinfo {note} {We point out that the energy transport in a
  temperature-biased JJ should include also another phase-dependent anomalous
  term, usually labelled as $P_{\protect \qopname \relax o{sin}}$ or
  $P_{\protect \text {pair}}$, which takes into account the heat transport
  originating from the energy-carrying tunneling processes involving Cooper
  pairs. Anyway, we stress that this term is a purely reactive
  contribution~\cite {Gol13,Vir17}, so that in the thermal balance
  equation~\protect \textup {\hbox {\mathsurround \z@ \protect \normalfont
  (\ignorespaces \ref {DynamicBalanceEqs}\unskip \@@italiccorr )}}, we have to
  neglect it.}\BibitemShut {Stop}%
      \bibitem [{\citenamefont {Kopnin}(2001)}]{Kop01}%
  \BibitemOpen
  \bibfield  {author} {\bibinfo {author} {\bibfnamefont {N.}~\bibnamefont
  {Kopnin}},\ }\href@noop {} {\emph {\bibinfo {title} {Theory of Nonequilibrium
  Superconductivity}}},\ International Series of Monogr\ (\bibinfo  {publisher}
  {Clarendon Oxford University Press},\ \bibinfo {year} {2001})\BibitemShut
  {NoStop}%
\bibitem [{\citenamefont {Timofeev}\ \emph {et~al.}(2009)\citenamefont
  {Timofeev}, \citenamefont {Garc\'{\i}a}, \citenamefont {Kopnin},
  \citenamefont {Savin}, \citenamefont {Meschke}, \citenamefont {Giazotto},\
  and\ \citenamefont {Pekola}}]{Pek09}%
  \BibitemOpen
  \bibfield  {author} {\bibinfo {author} {\bibfnamefont {A.~V.}\ \bibnamefont
  {Timofeev}}, \bibinfo {author} {\bibfnamefont {C.~Pascual}\ \bibnamefont
  {Garc\'{\i}a}}, \bibinfo {author} {\bibfnamefont {N.~B.}\ \bibnamefont
  {Kopnin}}, \bibinfo {author} {\bibfnamefont {A.~M.}\ \bibnamefont {Savin}},
  \bibinfo {author} {\bibfnamefont {M.}~\bibnamefont {Meschke}}, \bibinfo
  {author} {\bibfnamefont {F.}~\bibnamefont {Giazotto}}, \ and\ \bibinfo
  {author} {\bibfnamefont {J.~P.}\ \bibnamefont {Pekola}},\ }\bibfield  {title}
  {\enquote {\bibinfo {title} {Recombination-limited energy relaxation in a
  Bardeen-Cooper-Schrieffer superconductor},}\ }\href {\doibase
  10.1103/PhysRevLett.102.017003} {\bibfield  {journal} {\bibinfo  {journal}
  {Phys. Rev. Lett.}\ }\textbf {\bibinfo {volume} {102}},\ \bibinfo {pages}
  {017003} (\bibinfo {year} {2009})}\BibitemShut {NoStop}%
\bibitem [{\citenamefont {Bosisio}\ \emph {et~al.}(2016)\citenamefont
  {Bosisio}, \citenamefont {Solinas}, \citenamefont {Braggio},\ and\
  \citenamefont {Giazotto}}]{Bos16}%
  \BibitemOpen
  \bibfield  {author} {\bibinfo {author} {\bibfnamefont {R.}~\bibnamefont
  {Bosisio}}, \bibinfo {author} {\bibfnamefont {P.}~\bibnamefont {Solinas}},
  \bibinfo {author} {\bibfnamefont {A.}~\bibnamefont {Braggio}}, \ and\
  \bibinfo {author} {\bibfnamefont {F.}~\bibnamefont {Giazotto}},\ }\bibfield
  {title} {\enquote {\bibinfo {title} {Photonic heat conduction in
  Josephson-coupled Bardeen-Cooper-Schrieffer superconductors},}\ }\href
  {\doibase 10.1103/PhysRevB.93.144512} {\bibfield  {journal} {\bibinfo
  {journal} {Phys. Rev. B}\ }\textbf {\bibinfo {volume} {93}},\ \bibinfo
  {pages} {144512} (\bibinfo {year} {2016})}\BibitemShut {NoStop}%
\bibitem [{\citenamefont {Kraus}(1996)}]{Kra96}%
  \BibitemOpen
  \bibfield  {author} {\bibinfo {author} {\bibfnamefont {H.}~\bibnamefont
  {Kraus}},\ }\bibfield  {title} {\enquote {\bibinfo {title} {Superconductive
  bolometers and calorimeters},}\ }\href
  {http://stacks.iop.org/0953-2048/9/i=10/a=001} {\bibfield  {journal}
  {\bibinfo  {journal} {Supercond. Sci. Technol.}\ }\textbf {\bibinfo {volume}
  {9}},\ \bibinfo {pages} {827} (\bibinfo {year} {1996})}\BibitemShut {NoStop}%
\bibitem [{\citenamefont {Eisaman}\ \emph {et~al.}(2011)\citenamefont
  {Eisaman}, \citenamefont {Fan}, \citenamefont {Migdall},\ and\ \citenamefont
  {Polyakov}}]{Eis11}%
  \BibitemOpen
  \bibfield  {author} {\bibinfo {author} {\bibfnamefont {M.~D.}\ \bibnamefont
  {Eisaman}}, \bibinfo {author} {\bibfnamefont {J.}~\bibnamefont {Fan}},
  \bibinfo {author} {\bibfnamefont {A.}~\bibnamefont {Migdall}}, \ and\
  \bibinfo {author} {\bibfnamefont {S.~V.}\ \bibnamefont {Polyakov}},\
  }\bibfield  {title} {\enquote {\bibinfo {title} {Invited review article:
  Single--photon sources and detectors},}\ }\href {\doibase 10.1063/1.3610677}
  {\bibfield  {journal} {\bibinfo  {journal} {Rev. Sci. Instrum.}\ }\textbf
  {\bibinfo {volume} {82}},\ \bibinfo {pages} {071101} (\bibinfo {year}
  {2011})}\BibitemShut {NoStop}%
\bibitem [{\citenamefont {Berggren}\ \emph {et~al.}(2013)\citenamefont
  {Berggren}, \citenamefont {Dauler}, \citenamefont {Kerman}, \citenamefont
  {Nam},\ and\ \citenamefont {Rosenberg}}]{Ber13}%
  \BibitemOpen
  \bibfield  {author} {\bibinfo {author} {\bibfnamefont {K.~K.}\ \bibnamefont
  {Berggren}}, \bibinfo {author} {\bibfnamefont {E.~A.}\ \bibnamefont
  {Dauler}}, \bibinfo {author} {\bibfnamefont {A.~J.}\ \bibnamefont {Kerman}},
  \bibinfo {author} {\bibfnamefont {S.-W.}\ \bibnamefont {Nam}}, \ and\
  \bibinfo {author} {\bibfnamefont {D.}~\bibnamefont {Rosenberg}},\ }\bibfield
  {title} {\enquote {\bibinfo {title} {Detectors based on superconductors},}\
  }in\ \href@noop {} {\emph {\bibinfo {booktitle} {Experimental Methods in the
  Physical Sciences}}},\ Vol.~\bibinfo {volume} {45}\ (\bibinfo  {publisher}
  {Elsevier},\ \bibinfo {year} {2013})\ pp.\ \bibinfo {pages}
  {185--216}\BibitemShut {NoStop}%
\bibitem [{\citenamefont {Voutilainen}\ \emph {et~al.}(2010)\citenamefont
  {Voutilainen}, \citenamefont {Laakso},\ and\ \citenamefont
  {Heikkil\"a}}]{Vou10}%
  \BibitemOpen
  \bibfield  {author} {\bibinfo {author} {\bibfnamefont {J.}~\bibnamefont
  {Voutilainen}}, \bibinfo {author} {\bibfnamefont {M.~A.}\ \bibnamefont
  {Laakso}}, \ and\ \bibinfo {author} {\bibfnamefont {T.~T.}\ \bibnamefont
  {Heikkil\"a}},\ }\bibfield  {title} {\enquote {\bibinfo {title} {Physics of
  proximity Josephson sensor},}\ }\href {\doibase 10.1063/1.3354042} {\bibfield
   {journal} {\bibinfo  {journal} {J. Appl. Phys.}\ }\textbf {\bibinfo {volume}
  {107}},\ \bibinfo {pages} {064508} (\bibinfo {year} {2010})}\BibitemShut
  {NoStop}%
\bibitem [{\citenamefont {Virtanen}\ \emph {et~al.}(2018)\citenamefont
  {Virtanen}, \citenamefont {Ronzani},\ and\ \citenamefont {Giazotto}}]{Vir18}%
  \BibitemOpen
  \bibfield  {author} {\bibinfo {author} {\bibfnamefont {P.}~\bibnamefont
  {Virtanen}}, \bibinfo {author} {\bibfnamefont {A.}~\bibnamefont {Ronzani}}, \
  and\ \bibinfo {author} {\bibfnamefont {F.}~\bibnamefont {Giazotto}},\
  }\bibfield  {title} {\enquote {\bibinfo {title} {Josephson photodetectors via
  temperature-to-phase conversion},}\ }\href {\doibase
  10.1103/PhysRevApplied.9.054027} {\bibfield  {journal} {\bibinfo  {journal}
  {Phys. Rev. Applied}\ }\textbf {\bibinfo {volume} {9}},\ \bibinfo {pages}
  {054027} (\bibinfo {year} {2018})}\BibitemShut {NoStop}%
\bibitem [{\citenamefont {Kaplan}\ \emph {et~al.}(1976)\citenamefont {Kaplan},
  \citenamefont {Chi}, \citenamefont {Langenberg}, \citenamefont {Chang},
  \citenamefont {Jafarey},\ and\ \citenamefont {Scalapino}}]{Kap76}%
  \BibitemOpen
  \bibfield  {author} {\bibinfo {author} {\bibfnamefont {S.~B.}\ \bibnamefont
  {Kaplan}}, \bibinfo {author} {\bibfnamefont {C.~C.}\ \bibnamefont {Chi}},
  \bibinfo {author} {\bibfnamefont {D.~N.}\ \bibnamefont {Langenberg}},
  \bibinfo {author} {\bibfnamefont {J.~J.}\ \bibnamefont {Chang}}, \bibinfo
  {author} {\bibfnamefont {S.}~\bibnamefont {Jafarey}}, \ and\ \bibinfo
  {author} {\bibfnamefont {D.~J.}\ \bibnamefont {Scalapino}},\ }\bibfield
  {title} {\enquote {\bibinfo {title} {Quasiparticle and phonon lifetimes in
  superconductors},}\ }\href {\doibase 10.1103/PhysRevB.14.4854} {\bibfield
  {journal} {\bibinfo  {journal} {Phys. Rev. B}\ }\textbf {\bibinfo {volume}
  {14}},\ \bibinfo {pages} {4854--4873} (\bibinfo {year} {1976})}\BibitemShut
  {NoStop}%
\bibitem [{Note1()}]{Note1}%
  \BibitemOpen
  \bibinfo {note} {More specifically, the detector cannot reveal the arrival
  timing of a second photon, but it will anyway catch, due to the longer dead
  time, the additional energy associated to it.}\BibitemShut {Stop}%
\bibitem [{\citenamefont {Mates}\ \emph {et~al.}(2017)\citenamefont {Mates},
  \citenamefont {Becker}, \citenamefont {Bennett}, \citenamefont {Dober},
  \citenamefont {Gard}, \citenamefont {Hays-Wehle}, \citenamefont {Fowler},
  \citenamefont {Hilton}, \citenamefont {Reintsema}, \citenamefont {Schmidt},
  \citenamefont {Swetz}, \citenamefont {Vale},\ and\ \citenamefont
  {Ullom}}]{Mat17}%
  \BibitemOpen
  \bibfield  {author} {\bibinfo {author} {\bibfnamefont {J.~A.~B.}\
  \bibnamefont {Mates}}, \bibinfo {author} {\bibfnamefont {D.~T.}\ \bibnamefont
  {Becker}}, \bibinfo {author} {\bibfnamefont {D.~A.}\ \bibnamefont {Bennett}},
  \bibinfo {author} {\bibfnamefont {B.~J.}\ \bibnamefont {Dober}}, \bibinfo
  {author} {\bibfnamefont {J.~D.}\ \bibnamefont {Gard}}, \bibinfo {author}
  {\bibfnamefont {J.~P.}\ \bibnamefont {Hays-Wehle}}, \bibinfo {author}
  {\bibfnamefont {J.~W.}\ \bibnamefont {Fowler}}, \bibinfo {author}
  {\bibfnamefont {G.~C.}\ \bibnamefont {Hilton}}, \bibinfo {author}
  {\bibfnamefont {C.~D.}\ \bibnamefont {Reintsema}}, \bibinfo {author}
  {\bibfnamefont {D.~R.}\ \bibnamefont {Schmidt}}, \bibinfo {author}
  {\bibfnamefont {D.~S.}\ \bibnamefont {Swetz}}, \bibinfo {author}
  {\bibfnamefont {L.~R.}\ \bibnamefont {Vale}}, \ and\ \bibinfo {author}
  {\bibfnamefont {J.~N.}\ \bibnamefont {Ullom}},\ }\bibfield  {title} {\enquote
  {\bibinfo {title} {Simultaneous readout of 128 x-ray and gamma-ray
  transition--edge microcalorimeters using microwave SQUID multiplexing},}\
  }\href {\doibase 10.1063/1.4986222} {\bibfield  {journal} {\bibinfo
  {journal} {Appl. Phys. Lett.}\ }\textbf {\bibinfo {volume} {111}},\ \bibinfo
  {pages} {062601} (\bibinfo {year} {2017})}\BibitemShut {NoStop}%
\bibitem [{\citenamefont {Likharev}(1986)}]{Lik86}%
  \BibitemOpen
  \bibfield  {author} {\bibinfo {author} {\bibfnamefont {K.K.}\ \bibnamefont
  {Likharev}},\ }\href@noop {} {\emph {\bibinfo {title} {Dynamics of Josephson
  Junctions and Circuits}}}\ (\bibinfo  {publisher} {Gordon and Breach, New
  York},\ \bibinfo {year} {1986})\BibitemShut {NoStop}%
\bibitem [{\citenamefont {Govenius}\ \emph {et~al.}(2014)\citenamefont
  {Govenius}, \citenamefont {Lake}, \citenamefont {Tan}, \citenamefont
  {Pietil\"a}, \citenamefont {Julin}, \citenamefont {Maasilta}, \citenamefont
  {Virtanen},\ and\ \citenamefont {M\"ott\"onen}}]{Gov14}%
  \BibitemOpen
  \bibfield  {author} {\bibinfo {author} {\bibfnamefont {J.}~\bibnamefont
  {Govenius}}, \bibinfo {author} {\bibfnamefont {R.~E.}\ \bibnamefont {Lake}},
  \bibinfo {author} {\bibfnamefont {K.~Y.}\ \bibnamefont {Tan}}, \bibinfo
  {author} {\bibfnamefont {V.}~\bibnamefont {Pietil\"a}}, \bibinfo {author}
  {\bibfnamefont {J.~K.}\ \bibnamefont {Julin}}, \bibinfo {author}
  {\bibfnamefont {I.~J.}\ \bibnamefont {Maasilta}}, \bibinfo {author}
  {\bibfnamefont {P.}~\bibnamefont {Virtanen}}, \ and\ \bibinfo {author}
  {\bibfnamefont {M.}~\bibnamefont {M\"ott\"onen}},\ }\bibfield  {title}
  {\enquote {\bibinfo {title} {Microwave nanobolometer based on proximity
  Josephson junctions},}\ }\href {\doibase 10.1103/PhysRevB.90.064505}
  {\bibfield  {journal} {\bibinfo  {journal} {Phys. Rev. B}\ }\textbf {\bibinfo
  {volume} {90}},\ \bibinfo {pages} {064505} (\bibinfo {year}
  {2014})}\BibitemShut {NoStop}%
\bibitem [{\citenamefont {Govenius}\ \emph {et~al.}(2016)\citenamefont
  {Govenius}, \citenamefont {Lake}, \citenamefont {Tan},\ and\ \citenamefont
  {M\"ott\"onen}}]{Gov16}%
  \BibitemOpen
  \bibfield  {author} {\bibinfo {author} {\bibfnamefont {J.}~\bibnamefont
  {Govenius}}, \bibinfo {author} {\bibfnamefont {R.~E.}\ \bibnamefont {Lake}},
  \bibinfo {author} {\bibfnamefont {K.~Y.}\ \bibnamefont {Tan}}, \ and\
  \bibinfo {author} {\bibfnamefont {M.}~\bibnamefont {M\"ott\"onen}},\
  }\bibfield  {title} {\enquote {\bibinfo {title} {Detection of zepto Joule
  microwave pulses using electrothermal feedback in proximity-induced Josephson
  junctions},}\ }\href {\doibase 10.1103/PhysRevLett.117.030802} {\bibfield
  {journal} {\bibinfo  {journal} {Phys. Rev. Lett.}\ }\textbf {\bibinfo
  {volume} {117}},\ \bibinfo {pages} {030802} (\bibinfo {year}
  {2016})}\BibitemShut {NoStop}%
\bibitem [{\citenamefont {Day}\ \emph {et~al.}(2003)\citenamefont {Day},
  \citenamefont {LeDuc}, \citenamefont {Mazin}, \citenamefont {Vayonakis},\
  and\ \citenamefont {Zmuidzinas}}]{Day03}%
  \BibitemOpen
  \bibfield  {author} {\bibinfo {author} {\bibfnamefont {P.~K.}\ \bibnamefont
  {Day}}, \bibinfo {author} {\bibfnamefont {H.~G.}\ \bibnamefont {LeDuc}},
  \bibinfo {author} {\bibfnamefont {B.~A.}\ \bibnamefont {Mazin}}, \bibinfo
  {author} {\bibfnamefont {A.}~\bibnamefont {Vayonakis}}, \ and\ \bibinfo
  {author} {\bibfnamefont {J.}~\bibnamefont {Zmuidzinas}},\ }\bibfield  {title}
  {\enquote {\bibinfo {title} {A broadband superconducting detector suitable
  for use in large arrays},}\ }\href@noop {} {\bibfield  {journal} {\bibinfo
  {journal} {Nature}\ }\textbf {\bibinfo {volume} {425}},\ \bibinfo {pages}
  {817} (\bibinfo {year} {2003})}\BibitemShut {NoStop}%
\bibitem [{\citenamefont {Solinas}\ \emph {et~al.}(2018)\citenamefont
  {Solinas}, \citenamefont {Giazotto},\ and\ \citenamefont {Pepe}}]{Sol18}%
  \BibitemOpen
  \bibfield  {author} {\bibinfo {author} {\bibfnamefont {P.}~\bibnamefont
  {Solinas}}, \bibinfo {author} {\bibfnamefont {F.}~\bibnamefont {Giazotto}}, \
  and\ \bibinfo {author} {\bibfnamefont {G.~P.}\ \bibnamefont {Pepe}},\
  }\bibfield  {title} {\enquote {\bibinfo {title} {Proximity SQUID
  single-photon detector via temperature-to-voltage conversion},}\ }\href
  {\doibase 10.1103/PhysRevApplied.10.024015} {\bibfield  {journal} {\bibinfo
  {journal} {Phys. Rev. Applied}\ }\textbf {\bibinfo {volume} {10}},\ \bibinfo
  {pages} {024015} (\bibinfo {year} {2018})}\BibitemShut {NoStop}%
\bibitem [{\citenamefont {Guarcello}\ \emph
  {et~al.}(2018{\natexlab{c}})\citenamefont {Guarcello}, \citenamefont
  {Solinas}, \citenamefont {Braggio},\ and\ \citenamefont
  {Giazotto}}]{GuaSolBra18}%
  \BibitemOpen
  \bibfield  {author} {\bibinfo {author} {\bibfnamefont {C.}\ \bibnamefont
  {Guarcello}}, \bibinfo {author} {\bibfnamefont {P.}\ \bibnamefont
  {Solinas}}, \bibinfo {author} {\bibfnamefont {A.}\ \bibnamefont
  {Braggio}}, \ and\ \bibinfo {author} {\bibfnamefont {F.}\ \bibnamefont
  {Giazotto}},\ }\bibfield  {title} {\enquote {\bibinfo {title} {Solitonic
  thermal transport in a current-biased long Josephson junction},}\ }\href
  {\doibase 10.1103/PhysRevB.98.104501} {\bibfield  {journal} {\bibinfo
  {journal} {Phys. Rev. B}\ }\textbf {\bibinfo {volume} {98}},\ \bibinfo
  {pages} {104501} (\bibinfo {year} {2018}{\natexlab{c}})}\BibitemShut
  {NoStop}%
  \bibitem [{\citenamefont {Virtanen}\ \emph {et~al.}(2017)\citenamefont
  {Virtanen}, \citenamefont {Solinas},\ and\ \citenamefont {Giazotto}}]{Vir17}%
  \BibitemOpen
  \bibfield  {author} {\bibinfo {author} {\bibfnamefont {P.}~\bibnamefont
  {Virtanen}}, \bibinfo {author} {\bibfnamefont {P.}~\bibnamefont {Solinas}}, \
  and\ \bibinfo {author} {\bibfnamefont {F.}~\bibnamefont {Giazotto}},\ }\href
  {\doibase 10.1103/PhysRevB.95.144512} {\bibfield  {journal} {\bibinfo
  {journal} {Phys. Rev. B}\ }\textbf {\bibinfo {volume} {95}},\ \bibinfo
  {pages} {144512} (\bibinfo {year} {2017})}\BibitemShut {NoStop}%
\end{thebibliography}

%

\end{document}